\documentclass{aa}  
\usepackage{comment}
\usepackage{threeparttable}
\usepackage{caption}
\usepackage{booktabs}
\usepackage{subcaption}
\usepackage{tikz}
\usepackage{natbib}
\usepackage{amssymb}
\usepackage{graphicx}
\usepackage{hyperref}
\usepackage{float}
\usepackage{tabularx}
\usepackage{array}
\usepackage{siunitx}
\usepackage{textcomp}
\usepackage{txfonts}
\begin{document} 
   \title{The RoPES project with HARPS and HARPS-N }
   \subtitle{II. A third planet in the multi-planet system HD 176986}

   \author{N. Nari \inst{\ref{lbridges},\ref{iac},\ref{ull}}
\and
A. Su\'arez Mascare\~no \inst{\ref{iac},\ref{ull}}
\and
J. I. Gonz\'alez Hern\'andez \inst{\ref{iac},\ref{ull}}
\and
A. K. Stefanov \inst{\ref{iac},\ref{ull}}
\and
R. Rebolo \inst{\ref{iac},\ref{ull},\ref{csic}}
\and 
J. M. Mestre \inst{\ref{unipd}}
\and
X. Dumusque \inst{\ref{unige}}
\and
M. Cretignier \inst{\ref{unioxf}}
\and 
V. M. Passegger \inst{\ref{subaru},\ref{iac},\ref{ull},\ref{sw}}
\and
L. Mignon \inst{\ref{unigre},\ref{unige}}
}

   \institute{
Light Bridges S.L., Observatorio del Teide, Carretera del Observatorio, s/n Guimar, 38500, Tenerife, Canarias, Spain \label{lbridges}
\and
Instituto de Astrof{\'i}sica de Canarias, E-38205 La Laguna, Tenerife, Spain \label{iac}\\
\email{nicola.nari@lightbridges.es}
\and
Departamento de Astrof{\'i}sica, Universidad de La Laguna, E-38206 La Laguna, Tenerife, Spain \label{ull}
\and
Consejo Superior de Investigaciones Cient\'{i}ficas, Spain \label{csic}
\and
Dipartimento di Fisica e Astronomia "Galileo Galilei", Università di Padova, Vicolo dell’Osservatorio 3, 35122 Padova, Italy \label{unipd}
\and
Subaru Telescope, National Astronomical Observatory of Japan, 650 N Aohoku Place, Hilo, HI 96720, USA \label{subaru}
\and
Hamburger Sternwarte, Gojenbergsweg 112, 21029 Hamburg, Germany \label{sw}
\and
Observatoire de Gen\`eve, D\'epartement d'Astronomie, Universit\'e de Genève, Chemin Pegasi 51b, 1290 Versoix, Switzerland \label{unige}
\and 
Department of Physics, University of Oxford, Oxford OX13RH, UK \label{unioxf}
\and
Univ. Grenoble Alpes, CNRS, IPAG, F-38000 Grenoble, France \label{unigre}
}
 
  \abstract
   {Earth-like planets orbiting in the habitable zone of K- to G-type stars create an RV effect in amplitude of less than 1 \si{\meter\per\second} and have orbital periods of hundreds of days. Only long-term RV surveys with sub-meter per second precision instruments can explore the outer regions of Sun-like stars and look for Earth-like planets and super-Earths. Detecting Earth-like or super-Earth planets in the habitable zone of Sun-like stars is crucial to provide targets to the next generation of direct imaging facilities.}
   {We present the analysis of the K-type star HD 176986. It has a brightness of V=8.45 mag and a distance from the Sun of d = 27.88 pc. This star hosts a known planetary system of two super-Earths. We utilize historical and recently collected RV measurements to investigate the presence of Earth- and super-Earth-like planets in the habitable zone of HD 176986. 
   }
   {We monitored the system with HARPS and HARPS-N. We joined historical datasets with new data collected in an ongoing blind search program. We took advantage of recently developed tools for RV extraction and stellar activity filtering. The analysis of activity indicators permits us to determine the period of the magnetic cycle of the star alongside its rotation period. We performed a joint analysis of RVs and activity indicators through multidimensional GPs to better constrain the activity model in RVs and avoid overfitting. }
   {We detected a new planet orbiting the star and retrieved the two known planets. HD 176986 b has an orbital period of  6.49164$^{+0.00030}_{-0.00029}$ \si{\day} and a minimum mass of 5.36 $\pm$ 0.44 M$\oplus$. HD 176986 c has an orbital period of  P$_c$ = 16.8124 $\pm$ 0.0015 \si{\day} and a minimum mass of 9.75$_{-0.64}^{+0.65}$ M$\oplus$. HD 176986 d has an orbital period of 61.376$^{+0.051}_{-0.049}$ \si{\day} and a minimum mass of 6.76$_{-0.92}^{+0.91}$ M$\oplus$. From the analysis of activity indicators, we find evidence of a magnetic cycle with a period of 2432$_{-59}^{+64}$ \si{\day}, along with a rotation period of 36.05 $_{-0.71}^{+0.67}$\si{\day}.}
  {We discover a new planet in the multi-planet system orbiting the K-type star HD 176986. All the planets have minimum masses compatible with super-Earths or mini-Neptunes.} 

   \keywords{techniques: spectroscopic --
                techniques: radial velocities --
                planets and satellites: detection --
                stars: activity --
                planets and satellites: terrestrial planets --
                stars: individual: HD 176986
               }

   \maketitle
%

\section{Introduction}
The field of exoplanetary sciences has developed incessantly since the last decade of the 20th century. The first discovery of a planet orbiting around a main-sequence star \citep{1995_mayor_first_planet} paved the way for systematic research on exoplanets.
The latter finding was possible thanks to the radial velocity (RV) method \citep{struve_1952,hatzes_book}.  
Other techniques have also been applied to exoplanet detection through the years, such as the transit method \citep{2000_transit_charbonneau,2000_transit_henry}, which relies on the dimming of starlight when the planet passes in front of the star. 
The total number of exoplanets detected is rapidly increasing and has passed the number of 6000 confirmed planets up to November 27, 2025, reported in the \texttt{NASA Exoplanet Archive}\footnote{\url{https://exoplanetarchive.ipac.caltech.edu/}} \citep{2013_nasa_exoplanet_archive}.
Most of them were detected using the transit method. 

Even if outnumbered in terms of detections, the RV method remains fundamental for multiple reasons. This method is the most common technique for mass determination of exoplanets. The RV permits us to explore the outer regions of planetary systems, and long-term RV surveys are necessary to put well-determined constraints on the statistics of exoplanets at long orbital periods \citep{2011_mayor_occurrence,2022_pinamonti_occurrence,2024_bryan_occurrence_rate}. 
Furthermore, most of the planets in the solar neighborhood do not transit their star, and the RV method enables stars close to the Sun to be analyzed.

At first, the most precise instruments could reach a precision of 5-10 \si{\meter\per\second}, as was the case for ELODIE and HIRES \citep{1996_baranne_elodie,1994_hires_spectrograph}.
The High Accuracy Radial velocity Planet Searcher (HARPS; \citealp{2003_mayor_harps}), installed in 2003 at the Observatory of La Silla, Chile, brought a major improvement toward the detection of lighter planets. 
HARPS was the first spectrograph able to reach a precision on a single measurement of $<$ 1 \si{\meter\per\second}. The precision of HARPS was necessary for detecting new planetary mass ranges such as super-Earths \citep{2004_santos_SE}.  
Several years later, HARPS-N, an updated version of HARPS, was installed at the Telescopio Nazionale Galileo at the Observatory of La Palma \citep{2012_cosentino_harpsn,2014_harpsn_cosentino}. HARPS-N complements the coverage of the sky of HARPS due to the different latitudes of the two instruments.

In this context, the Rocky Planets in Equatorial Stars (RoPES) program \citep{2018_ropes_alejandro} is a long-term survey to investigate a sample of 17 G and K dwarf-type stars with HARPS and HARPS-N. The program aims to detect super-Earths in the habitable zone (HZ) of Sun-like stars and K-dwarfs. This task requires a time span of observations of several years. 
Detecting Earth and super-Earth like planets in the HZ of Sun-like stars will provide targets for future characterization of the atmosphere of exoplanets with the next generation of instruments such as ANDES \citep{2022_marconi_andes,2023_palle_andes} at the Extremely Large Telescope (ELT; \citealp{padovani_elt_2023}) and space missions such as Habitable World Observatory (HWO; \citealt{2024_mamajek_hwo}) and LIFE \citep{2022_life_quanz}.

HD 176986 is part of the RoPES program and is known for hosting a multi-planet system \citep{2018_ropes_alejandro}. 
This star has been observed through the years with HARPS and HARPS-N. With more than 300 nights of observations, it is the most intensively observed star of the program. 
The addition of new observations can bring further insights into the architecture of this multi-planet system. 
Furthermore, we applied newly developed tools in our analysis to mitigate systematic effects and stellar activity at the spectral level \citep{2021_yarara_cretignier,2023_yarara_cretignier}, and to model the activity of the star \citep{2015_rajpaul_multigp,2022_spleaf_delisle}.
\\
The work is structured as follows. In Sect. \ref{sec_obs}, we discuss the observations of the system used in our analysis. In Sect. \ref{sec_stellar_parameters}, we describe the star HD 176986. In Sect \ref{sec_analysis}, we describe in detail the analysis of RV and activity indicators time series. In Sect. \ref{sec_dis}, we discuss the findings of our analysis. In Sect. \ref{sec_con}, we summarize our work, and highlight the principal results. 
\section{Observations}
\label{sec_obs}
\subsection{HARPS}
HARPS \citep{2003_mayor_harps} is a high-resolution, fiber-fed spectrograph. It covers a wavelength range of between 380 \si{\nano\meter} and 690 \si{\nano\meter}. The spectrograph is a high-resolution spectrograph with a $\sim$ 115 000 resolving power. The instrument can reach an RV precision better than 50 \si{\centi\meter\per\second} on a single measurement \citep{2011_pepe_rv}. HARPS is contained in a vacuum vessel in a pressure- and temperature-stabilized environment to minimize spectral drifts due to pressure and temperature changes. 

The data reduction software (DRS) of the instrument provides wavelength-calibrated high-resolution spectra and RVs derived from the cross-correlation function (CCF) technique \citep{1953_fellgett_ccf}, alongside additional info as full width at half maximum (FWHM) and bisector time span (BIS). The CCF technique to extract velocities relies on the measurement of the Doppler displacement of the stellar spectrum relative to a stellar mask of a similar spectral type. For this work, we extracted RV and other data products with version 3.2.5 of the HARPS DRS, adapted from the DRS of ESPRESSO \footnote{\url{https:/www.eso.org/sci/software/pipe_aem_main.html}}. Version 3.2.5 of the DRS corrects for different systematics related to the ageing and change of lamps, among others.

HARPS underwent a fiber update in June 2015 \citep{2015_lo_curto_fiber}, which has improved the throughput of the instrument by 40 \% at 550 \si{\nano\meter}, as well as the instrument illumination and stability. The intervention introduced an offset compared to the RVs taken before it. This made it necessary to consider the observations taken before and after the fiber link update as two different datasets. We refer to HARPS dataset before the intervention as H03 and HARPS dataset after the intervention as H15. We fit a different zero-point and jitter for the H03 and H15 datasets. Details on the preparation of all the datasets used in our analysis are reported in Appendix \ref{methods_sec}. 
We collected 149 H03 spectra in 140 nights of observations and 246 H15 spectra in 119 nights of observations. For H03, the standard observing strategy was a single exposure of 900 \si{\second}. The median signal-to-noise ratio (S/N) reached was 133.8 at order 55. 
For H15, the standard observing strategy was to take three consecutive exposures of 300 \si{\second} each. We reached a median S/N of 71.0 at order 55.
\\
\subsection{HARPS-N}
HARPS-N \citep{2012_cosentino_harpsn} is a high-resolution, fiber-fed spectrograph. The instrument is based on the design of HARPS and it covers the spectral range between 380 \si{\nano\meter} and 690 \si{\nano\meter}. HARPS-N has a resolving power of $\sim$ 115 000 and its main goal is the detection of exoplanets in the Northern Hemisphere. The instrument is contained in a vacuum vessel. It is temperature- and pressure-stabilized in a similar way to HARPS. HARPS-N showed stability similar to HARPS, and a single measurement precision better than 50 \si{\centi\meter\per\second} \citep{2020_benatti_hd164922}. 

We refer to the HARPS-N datasets as HN. We collected 362 HN spectra in 109 nights. We performed three consecutive exposures of 300 \si{\second} each at every epoch as our standard observing strategy. The median S/N of the single exposures was 84.9 at order 55.  
\subsection{RV extraction}
In our analysis, we considered both RVs extracted with the CCF technique from the spectra derived by the DRS version 3.2.5 and RVs calculated by YARARA \citep{2021_yarara_cretignier,2023_yarara_cretignier} on the same spectra. YARARA is a tool to correct the spectra for systematics and extract RVs with a potential implementation of the line-by-line (LBL) technique \citep{2018_lbl_dumusque}. The LBL technique relies on a master spectrum built by merging all the spectra for a specific instrument. The following step is to measure the RV of every single line in the spectrum for all the observations. In this way, we obtain a time series of RVs of the single lines. Finally, the information coming from the single lines is merged to derive the RV related to each epoch. This method allows for the follow-up of the different behavior of the single lines, and to weight lines affected differently by stellar activity properly. 
The YARARA tool for the extraction of RVs applies a list of recipes to correct the spectra from systematics before extracting the RVs through the LBL or the CCF technique. These comprise correction for cosmic rays, interference patterns, telluric lines, point spread function variability, ghosts, stitching of the detector, and thorium-argon lamp contamination. 
YARARA can also correct stellar activity at the spectral level. This generates RVs and activity indicators cleaned for stellar activity. Depending on the needs of the analysis, we used both activity-corrected and non-activity-corrected datasets.
An exhaustive description of YARARA is given in \citet{2021_yarara_cretignier}. 
YARARA has a quality control on the spectra to discard spectra with problems related to S/N or anomalies in the CCF. Not all the nights of observations of HD 176986 have passed the quality control of YARARA. The YARARA dataset consists of 124 H03, 108 H15, and 98 HN nightly binned spectra, for a total dataset of 330 epochs. Only considering nights where the spectra passed the YARARA quality check, we can see a decrease in the root mean square (RMS) of the YARARA dataset compared to the dataset derived with the DRS. The RMS goes from 3.82 \si{\meter\per\second} to 3.68 \si{\meter\per\second} for the H03 dataset, from 3.67 \si{\meter\per\second} to 3.14 \si{\meter\per\second} for the H15 dataset, and from 3.51 \si{\meter\per\second} to 3.23 \si{\meter\per\second} for the HN dataset. The total RMS goes to from 3.68 \si{\meter\per\second} to 3.38 \si{\meter\per\second}. The mean error goes from 0.63 \si{\meter\per\second} to 0.42 \si{\meter\per\second} for H03, from 0.83 to 0.52 \si{\meter\per\second} for H15, and from 0.53 to 0.34 \si{\meter\per\second} for HN. We have also considered in the analysis the YARARA dataset without the stellar activity correction. In this way, we can assess the effectiveness of the YARARA correction. The YARARA dataset, not corrected for activity, shows an RMS of 3.51 \si{\meter\per\second} in the RV derived with the CCF technique. 
Where it is not stated differently, the YARARA dataset, not corrected for activity and with CCF-extracted RVs, has been used for the analysis of RVs and activity indicators. 
We show in Fig. \ref{hd176986_activity_indicators} the YARARA-RV and activity indicators time series. We show the RV and activity indicators not corrected for activity.

\subsection{TESS}
The Transiting Exoplanet Survey Satellite \citep[TESS;][]{tess_ricker_2014} observed HD 176986 in sector 80 with 120 \si{\second} and 20 \si{\second} cadence. We considered in our analysis the flux collected in bins of 120 \si{\second}. The data were processed through the Science Processing Operation Center (SPOC) pipeline \citep{2016_jenkins_spoc_tess}. The TESS mission provides different photometric light curves. Our analysis used the PDCSAP flux, automatically correcting artifacts and instrumental effects.  

We used a BLS periodogram \citep{2002_kovacs_bls} to investigate important periodicities present in the data. We did not find evidence of a transit. We refer to Appendix \ref{tess_appendix} for a brief analysis of the TESS dataset. We used the publicly available code \texttt{tpfplotter}
\footnote{\url{https://github.com/jlillo/tpfplotter}} \citep{2020_lillo_box_tpf_plotter} for plotting the target pixel file shown in the Fig. \ref{TPF_plot}. No additional sources within $\Delta$ Mag < 4 are found in the proximity of the target.
\subsection{ASAS-SN}

ASAS-SN is a photometric all-sky survey \citep{2014_shappee_asassn,2017_kochanek_asassn} developed to study supernovae and transients. The project is composed by 24 ground-based 14 \si{\centi\meter} telescopes part of the Los Cumbres Observatory Global Telescope Network. The telescopes are based in different sites worldwide, making it possible to survey the entire sky while minimizing the influence of local bad weather on the program. We collected a total of 3693 photometric measurements from ASAS-SN telescopes, 745 in the V band and 2948 in the g band. We only considered the measurement taken in the g-band. The analysis showed lunar contamination in our dataset and was not informative for the following analysis of RVs. We refer to Appendix \ref{asassn_appendix} for an analysis of the ASAS-SN dataset.
\section{HD 176986}
\label{sec_stellar_parameters}
\subsection{Stellar parameters}

HD 176986 is a K2.5 V star ($\alpha$ = 19:03:05.872 $\pm$ 0.020; $\delta$ = -11:02:38.131 $\pm$ 0.019; \citealt{2020_gaia_edr3}), close to the Solar System (d = 27.86 $\pm$ 0.02 pc; \citealt{2020_gaia_edr3}) and bright (V-mag = 8.45; \citealt{2000_tycho_catalogue}). 
In Table \ref{tab_stel_par} we present a summary of the main characteristics of the star.

The limits of the HZ depend on the luminosity of the star and its equilibrium temperature. In the case of HD 176986, we have L${\star}$ = 0.331 $\pm$ 0.027 L$\odot$ \citep{2008_sousa_lum} and T${\star}$ = 4931 $\pm$ 77 K \citep{2013_tsantaki_teff}. We followed the method of \citet{kopparapu_habitable_zone} and considered the case of a 1M$\oplus$ planet. They define four different regimes for calculating the HZ for a terrestrial planet, two to determine the inner edge of the HZ and two to infer the outer edge of the HZ. For an optimistic HZ, we considered their “recent Venus” and “early Mars” as inner and outer edges, respectively. For a conservative HZ, we considered their “runaway” and “maximum greenhouse” as inner and outer edges, respectively. We find the conservative HZ between 0.591 $\pm$ 0.024 AU and 1.046 $\pm$ 0.043 AU. This corresponds to an orbital period for a circular orbit between 186 $\pm$ 11 \si{\day} and 439 $\pm$ 28 \si{\day}. We find the optimistic HZ comprising between 0.446 $\pm$ 0.018 AU and 1.099 $\pm$ 0.045 AU. This corresponds to an orbital period for a circular orbit of between 122.5 $\pm$ 7.5 \si{\day} and 474 $\pm$ 30 \si{\day}.
\begin{table}[h!]
  \caption[]{Stellar parameters of interest for HD 176986.}
  \label{tab_stel_par}
  \begin{tabular}{p{0.5\linewidth}ll}
    \hline
    \hline
    \noalign{\smallskip}
    Parameter & HD 176986 & Ref\\
    \noalign{\smallskip}
    \hline
    \noalign{\smallskip}
    $\alpha$ & 19:03:05.872 & 1\\
    $\delta$ & -11:02:38.131 & 1\\
    Parallax (mas) & 35.858 $\pm$ 0.022 & 2 \\
    d (pc) & 27.86 $\pm$ 0.02 & 2 \\
    $\mu_{\alpha}$ $\cos$ $\delta$ (mas yr$^{-1}$) & -126.947 $\pm$ 0.023 & 2 \\
    $\mu_{\delta}$ (mas yr$^{-1}$) &  -235.938 $\pm$ 0.019 & 2 \\
    $T_\text{eff}$ (K) & $4931 \pm 77$ & 3 \\
    $\log_{10} g$ (cgs) & $4.44 \pm 0.17$ & 3 \\
    Spectral Type & K2.5V & 4\\
    RV (\si{\kilo\meter\per\second}) & 37.42 $\pm$ 0.14 & 2\\
    $[Fe/H]$ (dex) & 0.03 $\pm$ 0.05 & 3\\
    $M_{\star}$ (M$_{\odot}$) & 0.789 $\pm$ 0.019 & 3\\
    $R_{\star}$ (R$_{\odot}$) & 0.782 $\pm$ 0.035 & 5\\
    log$_{10}$R'$_{HK}$ & -4.90 $\pm$ 0.04 & 6 \\
    B (mag) & 9.39 $\pm$ 0.03 & 7 \\
    V (mag) & 8.45 $\pm$ 0.01 & 7 \\
    L (L$_{\odot}$) & 0.331 $\pm$ 0.027 & 8 \\
    Age (Gyr) & 4.3 $\pm$ 4.0 & 3 \\
    Inner optimistic HZ (AU) &  0.446 $\pm$ 0.018 & 0 \\
    Inner conservative HZ (AU) & 0.591 $\pm$ 0.024 & 0 \\
    Outer conservative HZ (AU) & 1.046 $\pm$ 0.043 & 0 \\
    Outer optimistic HZ (AU) & 1.099 $\pm$ 0.045 & 0 \\
    \noalign{\smallskip}
    \hline
  \end{tabular}
  \medskip 
    \begin{minipage}{0.5\textwidth}
        \raggedright
        References: 0 - This work, 1 - \citep{2020_edr3_gaia_catalogue}, 2 -\citep{2021_gaia_edr3}, 3 - \citep{2013_tsantaki_teff}, 4 - \citep{2006_gray_spectral_type}, 5 - \citep{2012_boyajian_diameter}, 6 -
        \citep{2018_ropes_alejandro}, 7 -
        \citep{2000_tycho_catalogue}, 8 - 
        \citep{2008_sousa_lum}.
    \end{minipage}
\end{table}
\subsection{Planetary system}

HD 176986 is a star known for hosting a multi-planet system. \citet{2018_ropes_alejandro} found two super-Earths orbiting the star. HD 176986 b has an orbital period of 6.4897 $\pm$ 0.0065 \si{\day} and a M$_b$$\sin$i$_b$ equal to 5.74 $\pm$ 0.66 M$\oplus$. HD 176986 c has an orbital period of 16.8191 $\pm$ 0.0044 \si{\day} and M$_c$$\sin$i$_c$ equal to 9.18 $\pm$ 0.97 M$\oplus$. A third significant signal was found at $\sim$ 35 \si{\day} but it was considered as related to stellar activity, as it was also present in activity indicators.

\section{Analysis}
\label{sec_analysis}

Here, we report the analysis of the HARPS and HARPS-N time series collected for HD 176986. To infer the parameters of the different models we used the nested sampling algorithm \citep{2004_skilling_nested_sampling} as it is implemented in Dynesty \citep{2020_destiny_speagle}. We performed a model comparison by comparing the evidence of the models applied to the same dataset. We refer to Appendix \ref{methods_sec} for details about the methods and tools used in our analysis to conduct parameter estimation and model comparison.
\subsection{Stellar activity}

Before moving to the analysis of the RVs, we analyzed the activity indicators at our disposal to retrieve information about the star, specifically the stellar rotation period and the period of the stellar magnetic cycle. We analyzed several activity indicators to characterize the stellar activity of HD 176986. We analyzed both CCF-derived metrics, such as the FWHM, bisector, and CCF contrast, and activity indicators related to the variability of specific lines of the spectra as Mount Wilson S index (S$_{MW}$; \citealt{1978_smw_vaughan}) and H$\alpha$ \citep{2011_gomes_da_silva_halpha}. 
A brief definition of the activity indicators we used in this work is outlined in Appendix \ref{stellar_activity_appendix}. 

We show in the panels on the left of Fig. \ref{hd176986_activity_indicators} the time series of the activity indicators and RVs derived by YARARA, without the correction for activity. We considered the time series not corrected for activity because in this way we can have a clearer signature of the magnetic cycle and the stellar rotation. In the right panels of Fig. \ref{hd176986_activity_indicators}, we show the GLS periodograms \citep{2009_gls_zeichmeister} of the RV and activity indicators time series. The GLS periodogram provides a preliminary indication of the most significant periodicities in our datasets, but it can be misleading if used to infer the best value for a parameter such as the magnetic cycle period or stellar rotation period. Still, the final values of the parameters are obtained in the global model. We explain our methods in Appendix \ref{methods_sec}.

We found the same period for the most prominent peak in FWHM, H$\alpha$, Contrast, and S index, at 2331 \si{\day}. We considered this period related to the magnetic cycle of the star, lying within the range of periods we would expect for a star of the spectral type of HD 176986 \citep{2011_lovis_cycle}. We see a peak at the same period with FAP $<$ 10 \% in the GLS periodogram of the Bisector, but in this case, the main peak in the periodogram is at 17.90 \si{\day}. We tried to model the cycle in the activity indicators with a sinusoid. We first considered for the analysis the S index, where we see the lower FAP value in correspondence of the 2331.89 \si{\day} peak. We used a wide uniform prior on the period $\mathcal{U}(1500d,3000d)$ to leave the model to choose the best value among a wide range of different periods. The period we found for the cycle is 2347 $\pm$ 33 \si{\day} for the S index. We used the same model for the other activity indicators. We found a result compatible with the result in the S index within 1$\sigma$ for all the other activity indicators. 

Once we recovered the magnetic cycle of the star, we tried to retrieve the rotation period of the star. We conducted a GP analysis using the GP package S+Leaf \citep{2020_spleaf_delisle,2022_spleaf_delisle}. We adopted the MEP kernel as a reference kernel in our analysis (Appendix \ref{stellar_activity_appendix}). We used a uniform prior, $\mathcal{U}(25d,45d)$, to fit for the rotation period. This prior encloses the period of the signal found in FWHM at $\sim$ 36 \si{\day} in \citet{2018_ropes_alejandro} and the period of the main peaks found in the periodogram of the residual of FWHM and S index, at 40.52 \si{\day} and 40.66 \si{\day}, respectively. The GP analysis gave us a converging result on the rotation period and the timescale of evolution of the activity pattern for S index, FWHM, Bisector, and Contrast, while we did not find a good result for H$\alpha$. The results we obtained for the rotation period are consistent among the different indicators at 1$\sigma$. We considered the value obtained in the S-index analysis as our reference value. We have a rotation period of 35.20 $^{+0.89}_{-0.71}$ \si{\day}. \citet{2015_suarez_mascareno_rhk} found the rotation period of the star to be 33.4 $\pm$ 0.2 \si{\day} from an analysis of chromospheric activity indicators as CA II H\&K and H$\alpha$. A signal at 35.74 \si{\day} was noticed in the RV analysis of \citet{2018_ropes_alejandro} and considered a signal related to activity.
The peak at 17.90 \si{\day} we see in the GLS periodogram of the bisector is at half the rotation period of the star. We concluded this peak is due to the first harmonic of the rotation-related activity signal. In Fig. \ref{fig_fwhm_stellar_activity} we show the activity model for FWHM as an example of the result we obtained. 

A similar analysis conducted on RVs revealed the presence of a weak signal related to the magnetic cycle, of amplitude K$_{cycle}$ = 0.87$_{-0.30}^{+0.29}$ \si{\meter\per\second} and period P$_{cycle}$ = 2439$_{-165}^{+238}$ \si{\day}, compatible with the result we found for the activity indicators.

We also ran a GP model on RVs, trying to recover the rotation period of the star, but it was not possible to find a converging result for the analysis of the RV dataset alone. The amplitude of the signal due to rotation is likely smaller than that of the planetary signals, and the GP struggles to model it effectively over the planets.

\subsection{Planetary signals research}

Once we had derived the rotation period and the period of the magnetic cycle, we proceeded to analyze the RV time series to search for planets. 
We had two different datasets at hand: YARARA-extracted RVs corrected from activity and systematics, and YARARA-extracted RVs only corrected for systematics. 
We conducted a parallel analysis to validate the results of each method. 
In the analysis of time series not corrected for activity, we took advantage of the analysis conducted on the activity indicators to guide the stellar activity model in the RVs. We considered the multidimensional GP framework \citep{2015_rajpaul_multigp,2023_barragan_multigp} to constrain the activity model, avoiding overfitting. 
A 3D GP, with RV and two activity indicators, has been shown to be a very effective model in stellar activity mitigation in \citet{2023_barragan_multigp}. Based on the results of the stellar activity analysis, we considered the two best activity indicators to use alongside RVs to be the FWHM and S index. We considered a term for the amplitude of the GP and the amplitude of the derivative of the GP in RVs and FWHM, while only a term for the amplitude of the GP in the S index.
In Sect. \ref{sec_dis} we compare the multidimensional approach with a 1D GP. 

\subsubsection{False inclusion probability}

\begin{figure*}[!htbp]
    \centering
    \includegraphics[width=\textwidth]{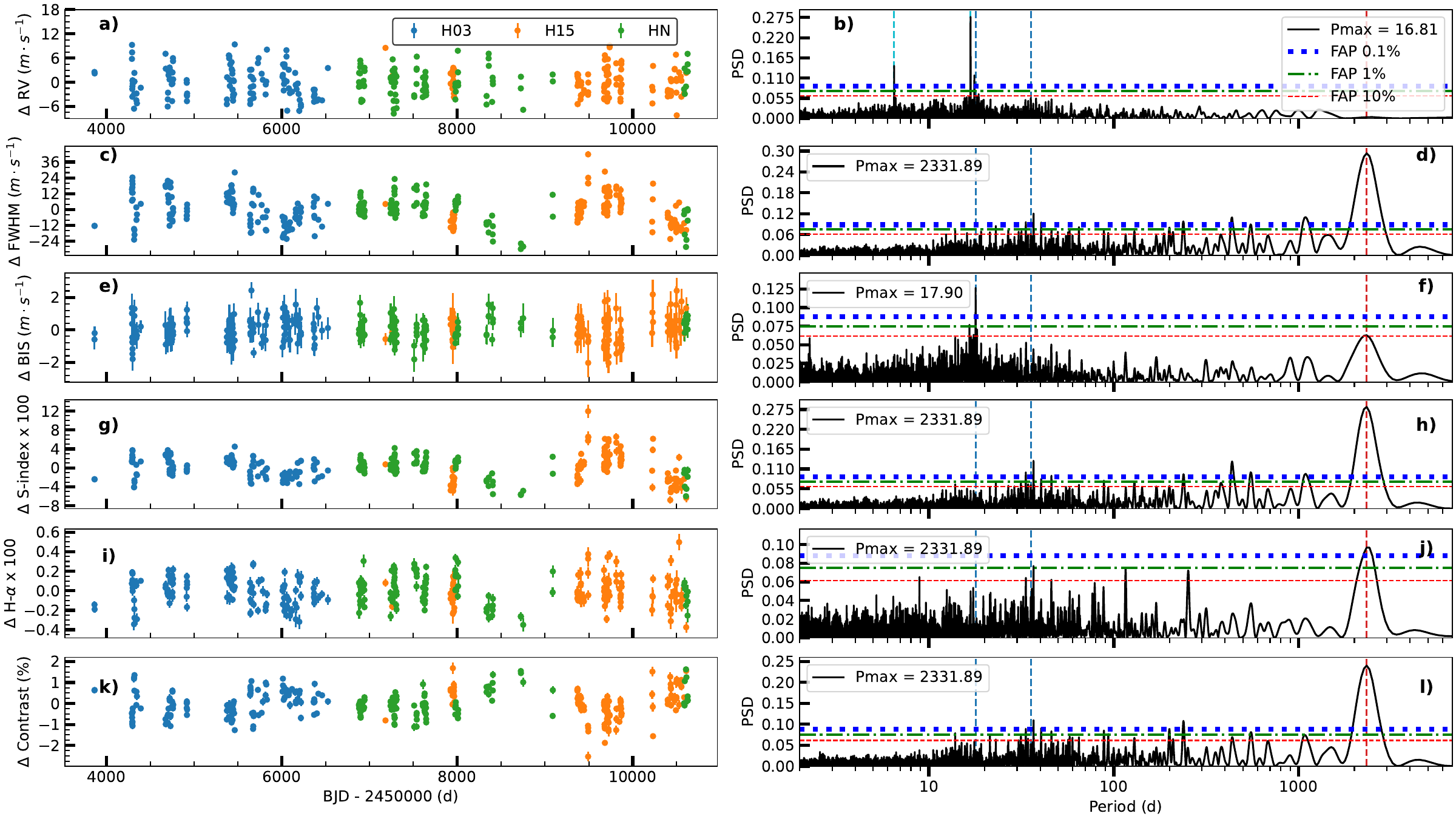}
    \caption{Time series of RV and activity indicators. Panel (a): RV time series. Panel (b): GLS periodogram related to the RV time series. We see two prominent peaks in correspondence with the two planets detected in \citet{2018_ropes_alejandro}. Panels (c), (e), (g), (i), and (k): Activity indicators time series. Panels (d), (f), (h), (j), and (l): Corresponding GLS periodograms. The red, green, and blue lines in the periodogram indicate a power corresponding to a 10\%, 1\%, and 0.1\% FAP, respectively. Panels (c) and (d): FWHM. We see a strong peak in the GLS periodogram at 2331.89 \si{\day}. Panels (e) and (f): Bisector. The strongest peak in the GLS periodogram is at 17.9 \si{\day} period. This peak is likely related to stellar rotation. A weaker peak at FAP $\sim$ 10 \% in correspondence with the magnetic cycle is also visible at a long period. Panels (g) and (h): S index. In panel (h) we see again a strong peak at 2331.89 \si{\day}, caused by the magnetic cycle. In panels (i) and (j), we see the time series and the GLS periodogram of H$\alpha$. We see again a peak at 2331.89 \si{\day} due to the magnetic cycle of the star. Panels (k) and (l): Time series and the GLS periodogram of Contrast. We can also find in this dataset the signature of the magnetic cycle in the peak at 2331.89 \si{\day} in the GLS periodogram. FWHM, S index, H$\alpha$, and Contrast all show an additional peak at $\sim$ 35 \si{\day}. This peak is related to the rotation period of the star as shown in the analysis of stellar activity.}
    \label{hd176986_activity_indicators}
\end{figure*}

We show in Fig. \ref{hd176986_activity_indicators}a the RV dataset. In Fig. \ref{hd176986_activity_indicators}b we show the GLS periodogram of the RV dataset. The RV dataset here was extracted with YARARA without stellar activity correction.

We found in the GLS periodogram significative peaks at $\sim$ 16.8 \si{\day} and $\sim$ 6.5 \si{\day}. These two peaks are compatible with the results of \citet{2018_ropes_alejandro}.

\begin{figure}[!h]
    \begin{minipage}{0.45\textwidth}
        \includegraphics[width=\linewidth]{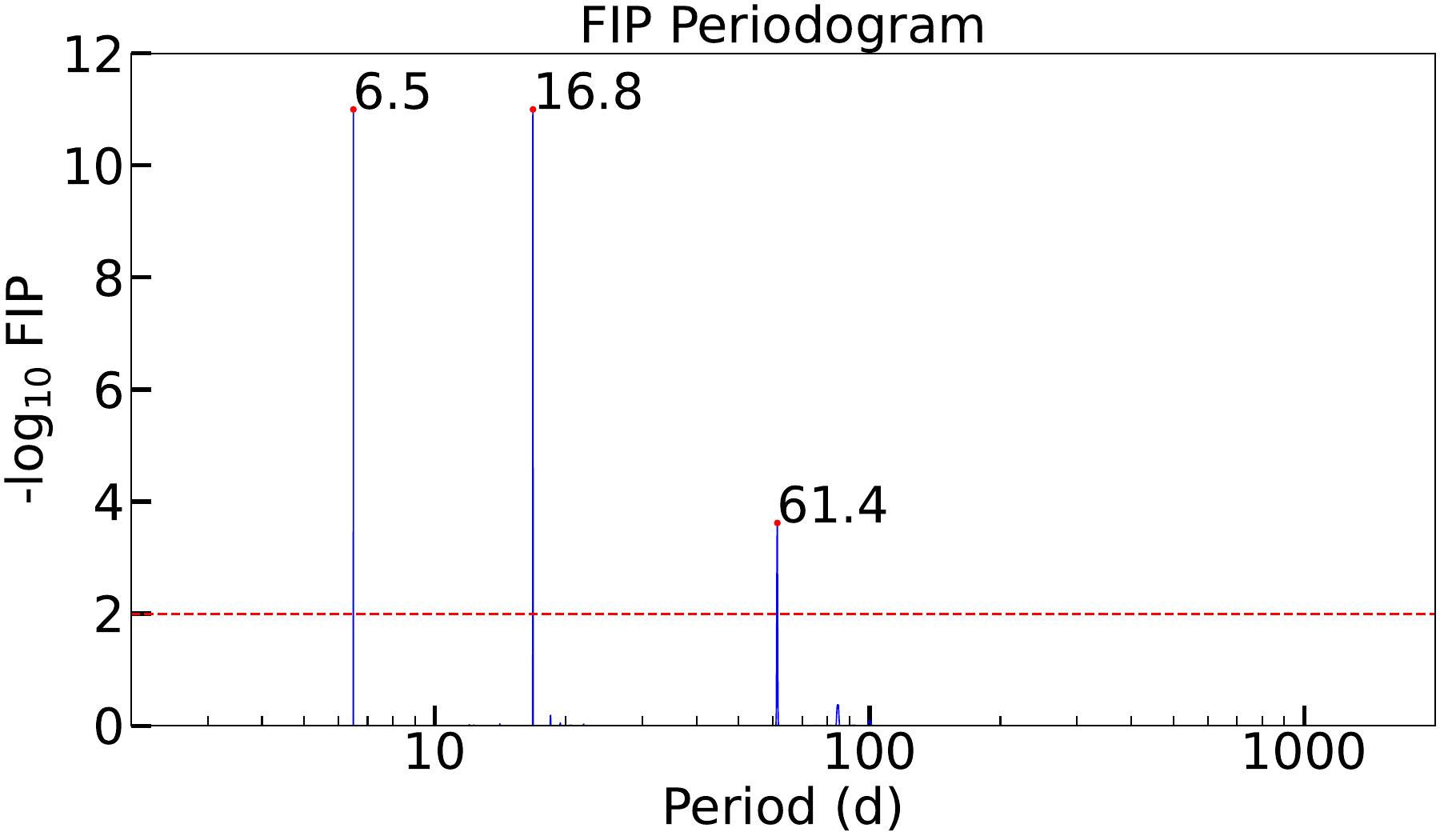}
    \end{minipage}
    \caption{FIP periodogram for HD 176986. The dashed red line represents the 1 \% probability threshold for the periodic signal to be a false positive. We see three peaks exceeding this threshold, at period of 6.5 \si{\day}, 16.8 \si{\day}, and 61.4 \si{\day}. The FIP level of 6.5 \si{\day} and 16.8 \si{\day} is cut at 11 because the numeric result gives an unphysical infinity. The peak at 61.4 \si{\day} was not retrieved in the previous analysis and is an original result of our work.
    }
    \label{hd176986_fip_periodogram}
\end{figure}

A more refined analysis for the inference of the significant periodic signals in a time series is defined in \citet{2022_fip_hara} through the definition of the false inclusion probability (FIP). 
The FIP is a metric we use to assess the significance of a signal of a determined period in a dataset by comparison to a reference value. The FIP can be computed using a correlated noise model, as is the case for GPs. We consider as signals of interest the signals with an amplitude in the FIP periodogram corresponding to a FIP < 1 \%. To have a reliable FIP analysis, we have to consider an effective model for stellar activity. We conducted a test considering the activity correction applied by YARARA, as well as a multidimensional GP modeling three time series: the S index, FWHM, and RV. We also put a cycle term for RV in the model. For the analysis with activity indicators, we considered the phase in the RV term of the magnetic cycle independent from the phase in the S index and FWHM, and we searched for it as a free parameter.
We ran a model with five sinusoids.
We show the FIP periodogram for the model where we considered the stellar activity corrected by YARARA in Fig. \ref{hd176986_fip_periodogram}.  

The FIP periodogram shows three peaks corresponding to periods of 6.5 \si{\day}, 16.8 \si{\day}, and 61.4 \si{\day}, all having a value of $-\log_{10} \text{FIP}$ above 2, which corresponds to a FIP $<$ 1 \%. 
The planets at 6.5 \si{\day} and the planet at 16.8 \si{\day} are already known from \citet{2018_ropes_alejandro}. The signal at 61.4 \si{\day} was not recovered in previous works on the system.
 
\subsubsection{Three-planet model}
 
We considered a model with three sinusoids and compared the evidence of this model with the evidence of a model with two sinusoids. To do this, we considered as a metric the logarithm of the evidence (lnZ) provided by Dynesty \citep{2020_destiny_speagle}. We considered a fully blind search for the planets. For the two-planet model, we considered two sinusoids with a log-uniform prior on the period between 1 \si{\day} and 2000 \si{\day}. We used the same priors on the period of the three planets also for the three-planet model. We tested both the activity-corrected and the non-activity-corrected YARARA dataset. For the activity-corrected dataset, we considered an RV-only model, relying on the correction of activity made by YARARA. In the case of the non-activity-corrected dataset, we worked in a multidimensional GP framework where we modeled the FWHM and S index alongside RVs. In both cases, we found an improvement in the logarithm of the evidence of the model of more than 5. This is the threshold we used to favor a more complex model over a simpler model (Appendix \ref{methods_sec}). For the YARARA-corrected activity dataset, we found a $\Delta$ lnZ = + 16.1. We found a signal amplitude of K = 1.19 $_{-0.17}^{+0.16}$ \si{\meter\per\second} and a period of P = 61.363$\pm$ 0.048 \si{\day}. In the case of the analysis with multidimensional GP, we recover a more significant result, with an improvement $\Delta$ lnZ = +23.6 after the addition of the new signal. In this case, the amplitude of the signal is K = 1.29 $^{+0.18}_{-0.17}$ \si{\meter\per\second} and the orbital period is recovered to be 61.378$\pm$ 0.51.
We recover the signal at 61.4 \si{\day} with two different approaches with a similar confidence. This shows the effectiveness of the stellar activity correction by YARARA. 
We tested a four-sinusoid model. Compared to the three-planet model, the four-planet model yields $\Delta$ lnZ = -1.5 for the activity-corrected dataset and $\Delta$ lnZ = -10.3 for the non-activity-corrected dataset. We were not able to find a period of the hypothetical fourth planet in the activity corrected dataset. In the non-activity corrected dataset, we see a peak in the posterior distribution of the period at $\sim$ 215 \si{\day}, but the four-planet model is strongly disfavored in terms of model evidence. We adopted the three-planet model as our reference model based on the Bayesian evidence. 
\subsubsection{Keplerian signals}

We tried to model the three signals we detected in our analysis with a Keplerian function instead of a sinusoidal one. We parametrized a combination of the eccentricity, e, and the argument of the periastron, $\omega$, $\sqrt{e} \cdot \cos(\omega)$ and $\sqrt{e} \cdot \sin(\omega)$ \citep{2011_anderson_eccentricity,2013_eastman_eccentricity}.
We used a normal prior for the parameter: $\mathcal{N}$[0,0.3], with the constraint, expressed inside the likelihood, to have e < 0.99. This avoids the sampler from exploring an unphysical region of the parameter space.
We considered the multidimensional GP model with the time series not corrected for activity. 
We were only able to put upper limits on the eccentricity of the planets. We find for HD 176986 b an upper limit on the eccentricity of e$_b$ $<$ 0.08 and for HD 176986 an upper limit of e$_c$ $<$ 0.11. Also, for the new signal at 61.4 \si{\day} we are only able to find an upper limit of e$_d$ $<$ 0.31.
To calculate the upper limit, we considered the 84th percentile of the posterior distribution of the parameter. We consider this value to be consistent with the calculation of the positive error of the measurements of the other parameters.
The model with three Keplerians is only favored with respect to the model with three sinusoids by a $\Delta$ lnZ = +0.2. This, in conjunction with the lack of a significant result on the eccentricity, led us to adopt the sinusoidal model as our reference model. 

The same test was conducted with the dataset corrected by YARARA for stellar activity. We obtained an upper limit of e$_b$ $<$ 0.11 for HD 176986 b, e$_c$ $<$ 0.08 for HD 176986 c, and e$_d$ $<$ 0.61 for the new signal at 61.4 \si{\day}. The result for the outer planet is less defined than in the case of the multidimensional GP. Anyway, the error in the calculation prevents us from retrieving an eccentricity value at a high confidence level. The eccentric model is disfavored in lnZ by $\Delta$ lnZ = -6. This confirms the preference for the model with circular orbits over the model with eccentric orbits. The problem of an overestimation of the eccentricities of planets was considered in \citet{2019_hara_eccentricity}. The absence of a correction for the correlated noise, for example, the noise due to stellar activity, can lead the model to fit for a spurious eccentricity. In our case, we were considering the correction for stellar activity made by YARARA as our red-noise model, without further correction. We considered the upper limit in eccentricity larger than 0.50 due to some remnants in the activity, not completely ruled out by YARARA. 

\subsubsection{Model comparison}

We summarize in Table \ref{table_lnz} the results on the evidence of the models. We report the difference in lnZ of the two-sinusoid, three-sinusoid, four-sinusoid, and three-Keplerian models. We considered both the dataset with the activity corrected by YARARA and the dataset for which we applied a multidimensional GP. 
In both datasets, we took the three-sinusoid model as the zero-point for the $\Delta$ lnZ analysis because this was the best model in our analysis. 
We considered a uniform prior on the period of HD 176986 b $\mathcal{U}$[5d,10d], HD 176986 c $\mathcal{U}$[10d,20d], and $\mathcal{U}$[50d,70d] to refine the planetary ephemeris. The results are reported in Sect. \ref{planetary_system_disc}.
\begin{table}[h!]
  \centering
  \begin{threeparttable}
    \caption{Evidence of different models for HD 176986.} 
    \label{table_lnz}
    \begin{tabular}{l r r}
    \hline
    \hline
    \noalign{\smallskip}
    Model & $\Delta$ lnZ YARARA & $\Delta$ lnZ multi-dim GP \\
    \noalign{\smallskip}
    \hline
    \noalign{\smallskip}
      Two-sinusoidal    & $-16.1$ & $-23.6$ \\
      Three-sinusoidal  & $0.0$   & $0.0$   \\
      Three Keplerian   & $-6.0$  & $+0.2$  \\
      Four-sinusoidal   & $-1.5$  & $-10.3$  \\
      \bottomrule
    \end{tabular}
    \medskip 
    \begin{minipage}{0.5\textwidth}
        \raggedright
        Notes: We see in the table the difference in evidence of different models. The three-sinusoidal model is favored compared to the two-sinusoidal model and the four-sinusoidal model. The circular solution is also favored compared to the solution with an eccentricity.
    \end{minipage}
  \end{threeparttable}
\end{table}
\subsubsection{Stability of the signal}
\label{sec_apodized}
After the evidence of a signal at an orbital period of 61.4 \si{\day}, we tested its stability. Planetary signals are stable at every epoch, while activity-related signals can vary in amplitude with time \citep{2014_robertson_gj581,2021_lubin_barnard}. We tested if the signal at 61.4 \si{\day} was present at some specific epoch only.
This approach was inspired by the work of \citet{2022_hara_apodized}. We refer to it as the apodized test, reconnecting to the name of the test in \citet{2022_hara_apodized}.
We considered a model in which the planetary signal of interest is a sinusoid multiplied by a Gaussian with a center, $\mu$, and a width, $\sigma$, as is shown in Eq. \ref{apodized_equation}. The center of the Gaussian represents the location of the signal in the time series and the width is the timescale on which the signal is constant. If the signal is of planetary origin, it should not be located in a specific subset of the time series. Due to this, we should find a value for $\mu$ compatible with any point in the baseline and a large value for $\sigma$ if the signal was persistent throughout the entire time series.

\begin{equation}
    -K \cdot \sin\left(2\pi \cdot \frac{t - t_0}{P_{\text{pl}}}\right) \cdot G(\mu, \sigma)
    \label{apodized_equation}
\end{equation}

For this test, we considered our best model (three sinusoids). We considered both the dataset corrected for the activity and the multidimensional GP framework, and we found no relevant differences. We present the results of the multidimensional GP model for brevity.
 
 We repeated the test for the signals at 6.5 \si{\day}, 16.8 \si{\day}, and 61.4 \si{\day}. For all the tests, we used a uniform prior on the center of the Gaussian, $\mu$ \si{\day} $\mathcal{U}$[2450000 BJD;2462000 BJD]. For the width of the Gaussian, $\sigma$, we considered a prior in log space $\mathcal{LU}$[0;20]. For the signal at 61.4 \si{\day} we obtained $\mu$ = 2455700 $^{+4000}_{-3800}$ BJD and ln $\sigma$ = 13.61 $^{+4.34}_{-4.17}$.

\begin{figure}[htbp]
    \begin{subfigure}[t]{0.48\textwidth}
        \centering
        \begin{tikzpicture}
            \node[anchor=north west, inner sep=0] (image) at (0,0) {\includegraphics[width=\linewidth]{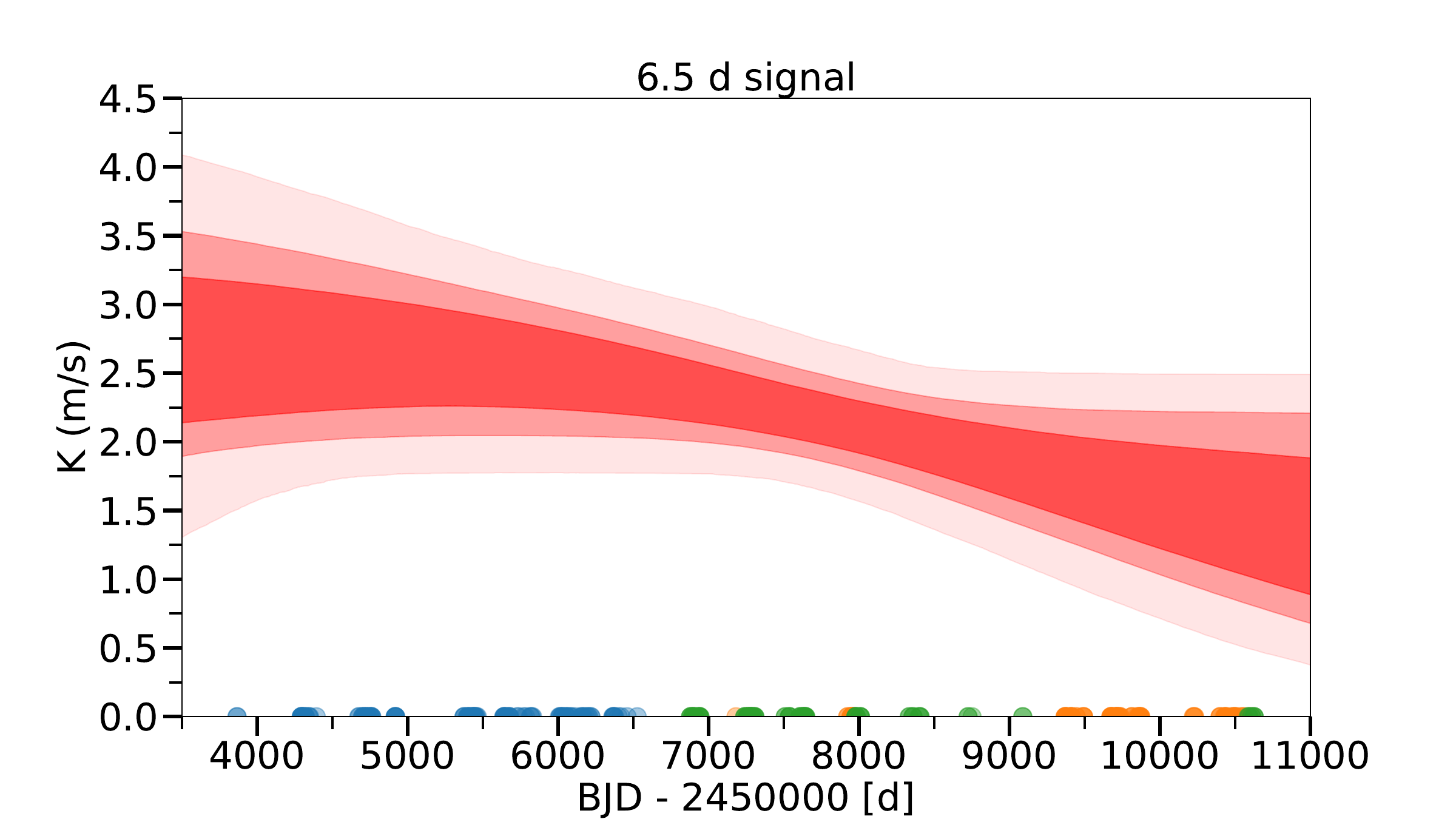}};
            \node[anchor=north west, inner sep=0] at (0.3, -0.6) {(a)}; 
        \end{tikzpicture}
        \label{fig:lower} 
    \end{subfigure}
    \hfill
    \begin{subfigure}[t]{0.48\textwidth}
        \centering
        \begin{tikzpicture}
            \node[anchor=north west, inner sep=0] (image) at (0,0) {\includegraphics[width=\linewidth]{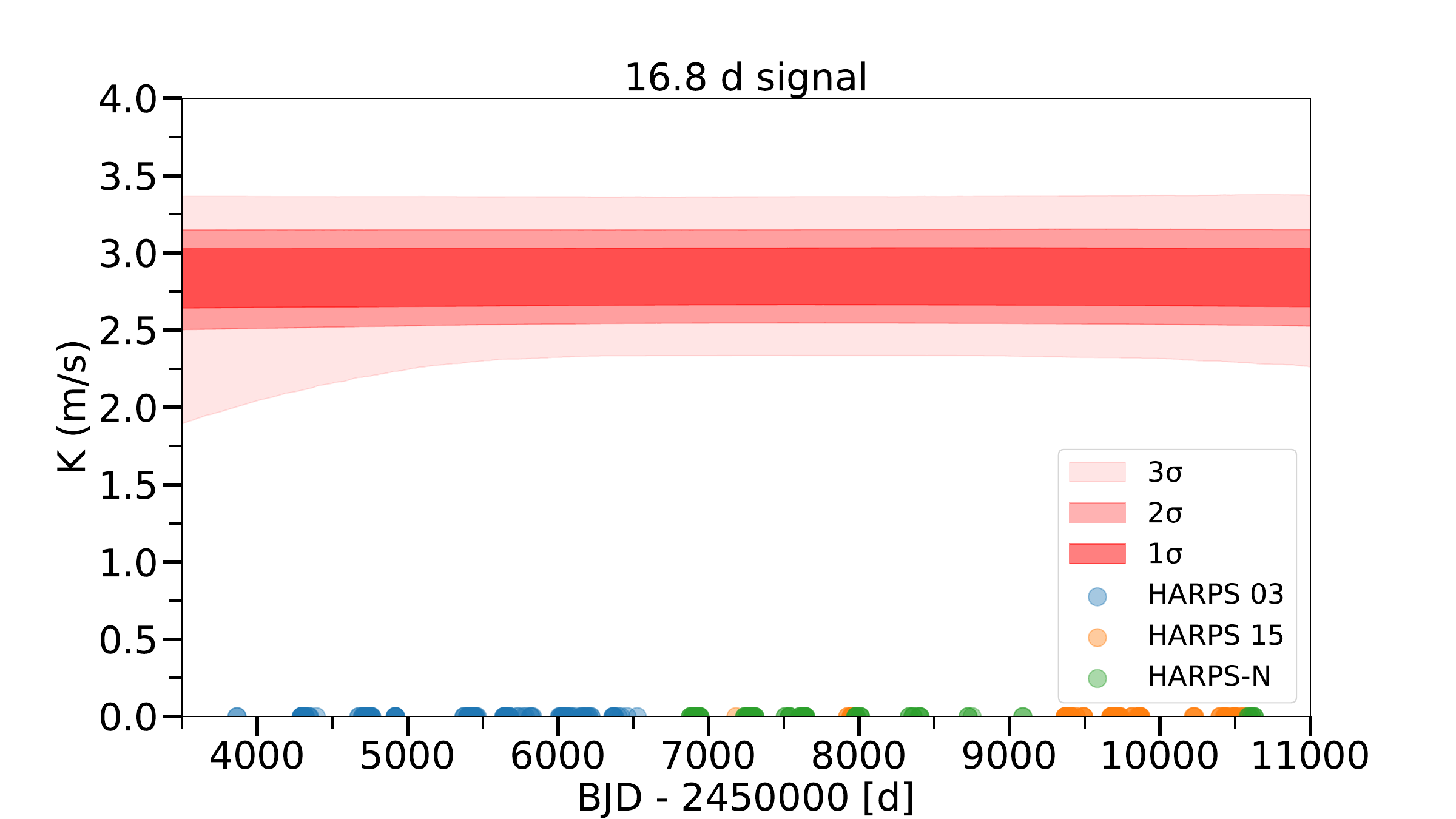}};
            \node[anchor=north west, inner sep=0] at (0.3, -0.6) {(b)}; 
        \end{tikzpicture}
        \label{fig:lower} 
    \end{subfigure}
    \hfill
    \centering
    \begin{subfigure}[t]{0.48\textwidth}
        \centering
        \begin{tikzpicture}
            \node[anchor=north west, inner sep=0] (image) at (0,0) {\includegraphics[width=\linewidth]{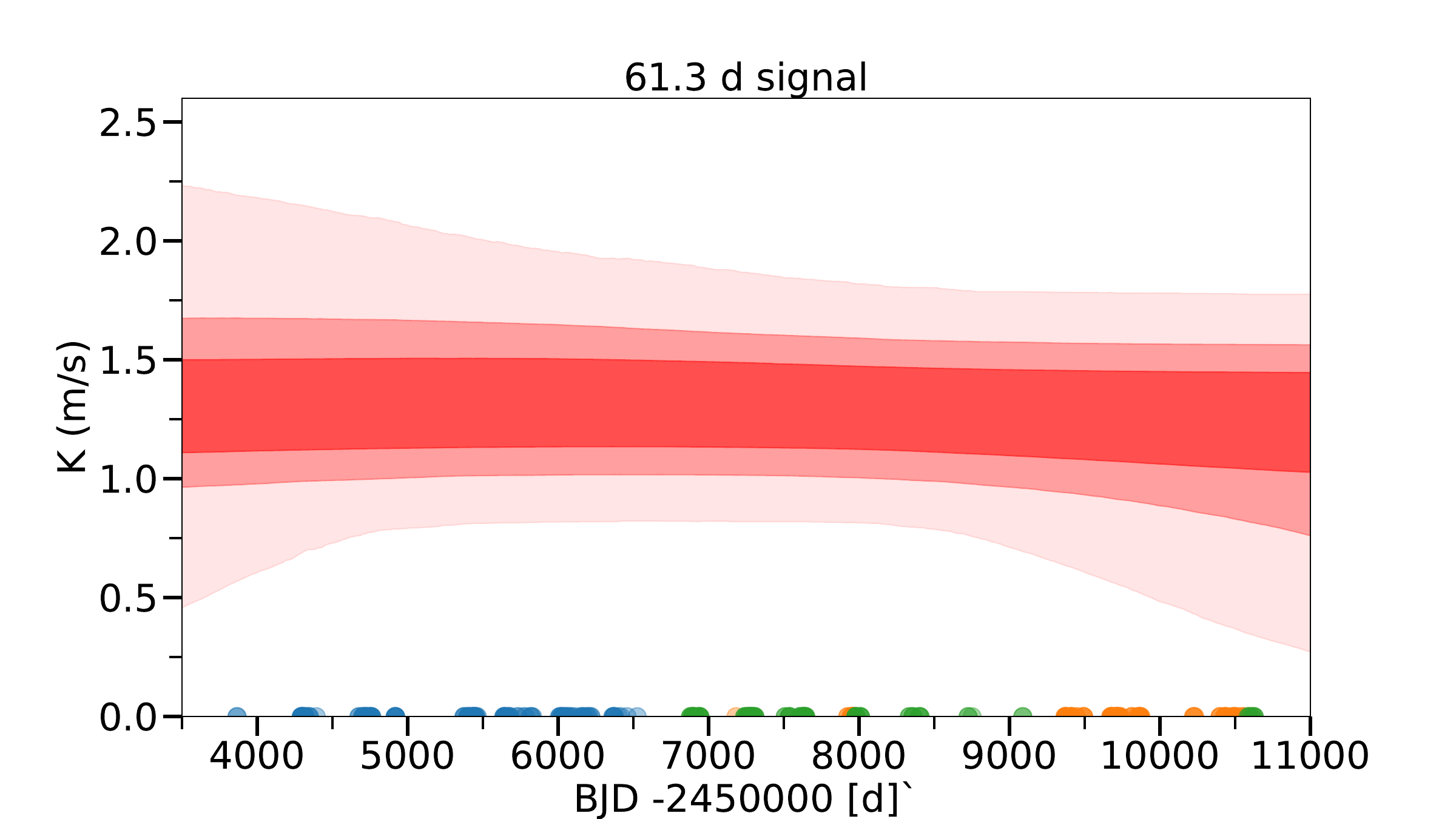}};
            \node[anchor=north west, inner sep=0] at (0.3, -0.6) {(c)}; 
        \end{tikzpicture}
        \label{fig:upper} 
    \end{subfigure}
    \caption{Apodized test for the three planets orbiting around HD 176986. Color-coded dots represent the epochs of observations of different instruments. Panel (a): Apodized signal for the 6.5 \si{\day} planet found in \citet{2018_ropes_alejandro}. The signal is stable throughout the time series, but it shows a different amplitude at different epochs. Panel (b): Apodized signal for the 16.8 \si{\day} planet. The signal is present in the full dataset and it is stable over time. 
    Panel (c): Apodized signal for the 61.4 \si{\day} candidate. The signal is stable over time.}
    \label{fig_apodized}
\end{figure}
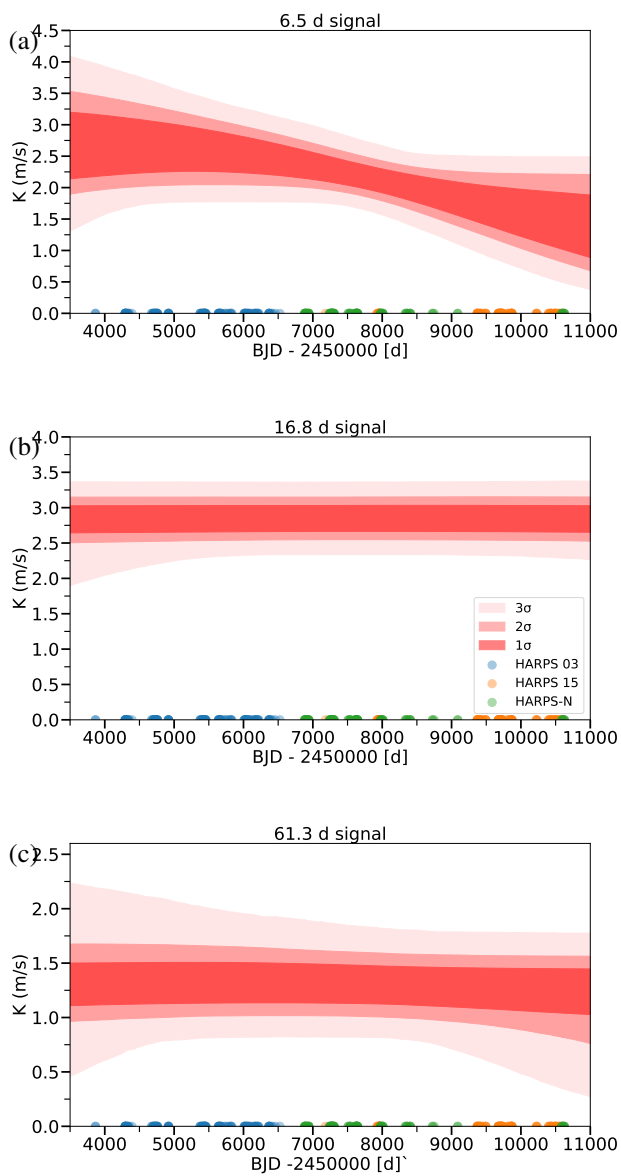

We show in Fig. \ref{fig_apodized} the product of the posterior distribution of the planet multiplied by the Gaussian filter. 
The signal is stable over the different epochs. 

We applied the same test for the two planets announced in \citet{2018_ropes_alejandro}. For HD 176986c, we find the signal to be stable over the observing seasons. We obtained for this planet $\mu$ = 6300 $^{+3900}_{-4200}$ and ln $\sigma$ = 14.6 $\pm$ 3.7. This is strong evidence in the direction of the stability of the signal, because the width of the Gaussian is $\sim$ a factor 100 larger in magnitude than the timescale of observations. 
For HD 176986 b, we found a less stable behavior. In this case we have $\mu$ = 2453500 $^{+2200}_{-2400}$ BJD and ln $\sigma$ = 8.74 $^{0.71}_{-0.40}$. The center of the Gaussian is still not strictly defined, but the model prefers to localize the signal at epochs with BJD $<$ 2456000. The width of the Gaussian is comparable to the time span of the observations. We see by the analysis of the posterior of the apodized amplitude in Fig. \ref{fig_apodized} how the amplitude of the signal decreases with time, being stronger in amplitude in the part of the dataset at BJD $<$ 2456000, but does not disappear even at the end of the time series, with an inflection point around 2458000 BJD. We conducted a series of tests on the presence of the planet in the different datasets. We tested whether the signal was present only in one of the three instruments: H03, H15, and HN. We found the signal in all three datasets, but with a different amplitude. For H03, we found K$_b$ = 2.72 $\pm$ 0.33 \si{\meter\per\second}. For H15, we found K$_b$ = 1.42 $_{-0.30}^{+0.28}$ \si{\meter\per\second}. For HN we found K$_b$ = 2.08 $\pm$ 0.27 \si{\meter\per\second}.
We see the amplitude value varies at more than 1$\sigma$ among the datasets, but all the signals are defined with a significance larger than 4$\sigma$. 
We also explored the phase stability of the signal. 
We calculated the time of conjunction using a common prior for the model of the three instruments: $\mathcal{U}[2456500-2456506.5]$. We found T0$_b$ = 6502.17 $\pm$ 0.23 BJD-2450000 for H03, T0$_b$ = 6501.56 $^{+0.71}_{-0.67}$ BJD-2450000 for H15, and T0$_b$ = 6501.60 $\pm$ 0.22 BJD-2450000 for HN.

The difference in amplitude could be the reason why we see the apodized amplitude change throughout the time series. The presence of the signal in all the datasets indicates that the signal is not due to instrumental systematics, being present in different instruments. 

We tried to recover the signal using the RV as extracted from the DRS-CCF instead of YARARA. In this case, we found more homogeneity in the amplitude when comparing the models of different instruments. For H03 we obtained K$_b$ = 2.34 $\pm$ 0.22 \si{\meter\per\second}, for H15 we obtained K$_b$ = 1.70 $_{-0.26}^{+0.25}$ \si{\meter\per\second}, and for HN we obtained  K$_b$ = 2.55 $_{-0.30}^{+0.29}$ \si{\meter\per\second}.

\begin{figure}[!h]
    \begin{minipage}{0.45\textwidth}
        \includegraphics[width=\linewidth]{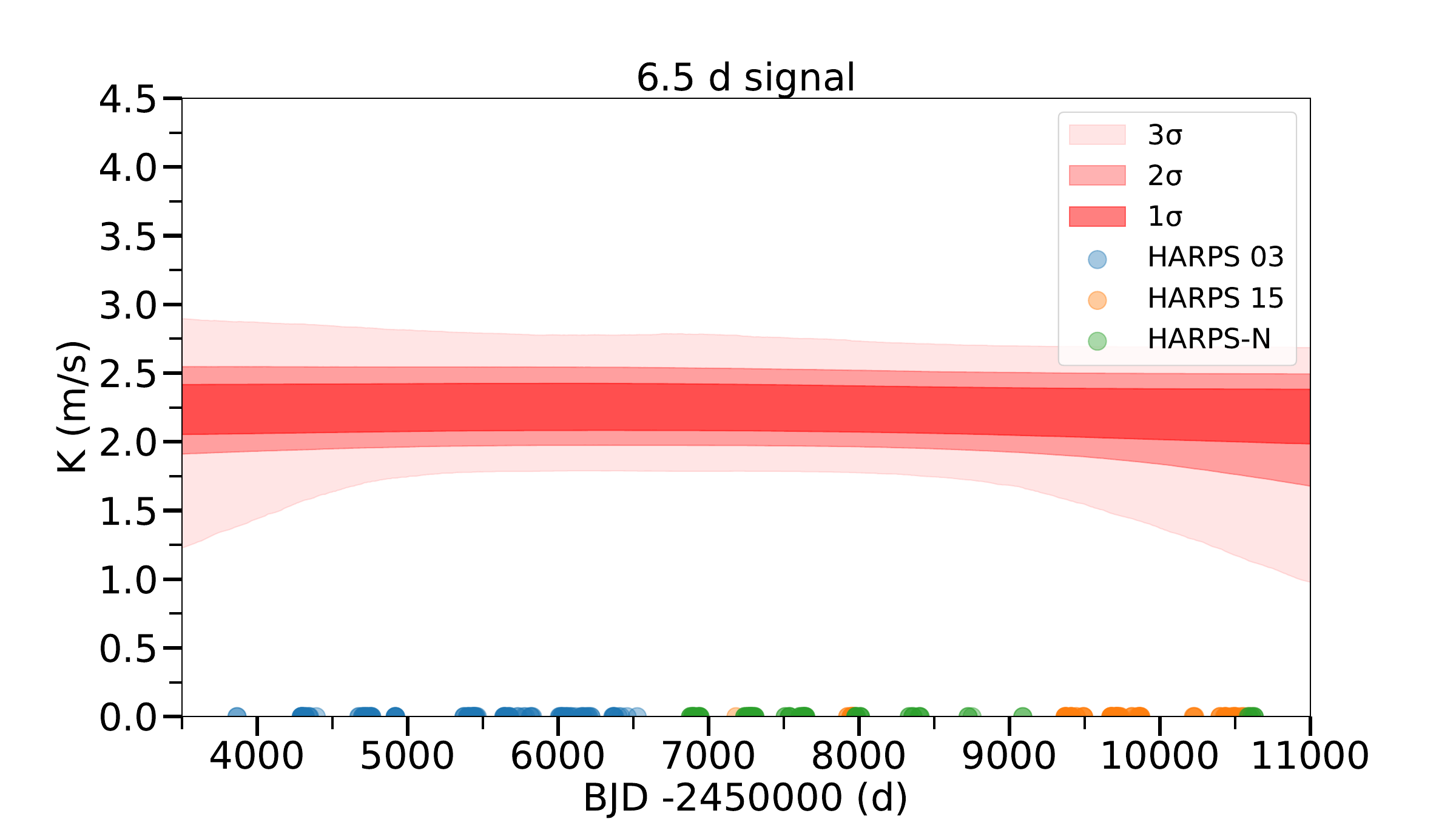}
    \end{minipage}
    \caption{Apodized test for the 6\si{\day} signal with the CCF RVs. The signal is stable throughout the full time series and it is not localized in any specific epoch.}
    \label{apodized_ccf}
\end{figure}

We also computed the apodized test for the 6.5 \si{\day} signal with the DRS-CCF dataset. In this case, we obtained a result in line with a signal persistent throughout the full dataset. We found $\mu$ = 2455900 $\pm$ 3900 BJD and ln $\sigma$ = 13.9$_{-4.1}^{+4.2}$. In Fig. \ref{apodized_ccf} we show the apodized amplitude of the signal.

We also performed a different test, in which we multiplied the sinusoid by a box function, B($\mu$,$\sigma$). 
The function, \( B(\mu, \sigma) \), is defined in Eq. \ref{box_equation}:
\begin{equation}
B(\mu, \sigma) = 
\begin{cases} 
1 & \text{if } \mu - \sigma < x < \mu + \sigma, \\
0 & \text{otherwise.}
\end{cases}
\label{box_equation}
\end{equation}

We repeated the test for the three planets of the system using the dataset with the stellar activity correction by YARARA to minimize the computational time of the test. 
A plot of the box amplitude is shown in Fig. \ref{fig_box}
We found for HD 176986 b $\mu_{b}$ = 2455500 $\pm$ 3500 BJD and log $\sigma_{b}$ = 10.4$_{-1.8}^{+3.2}$.
For HD 176986 c we found $\mu_{b}$ = 2456800 $_{-4200}^{+3800}$ 3500 BJD and log $\sigma_{b}$ = 11.9$\pm$ 2.1. For the new discovered HD 176986 d we found $\mu_{b}$ = 2456200 $\pm$ $_{-4000}^{+3900}$ BJD and log $\sigma_{b}$ = 11.8$_{-2.2}^{+2.1}$. 

We see the planet at 6.5 \si{\day} as the planet spanning the lower width of the box, but, as it is possible to see in Fig. \ref{fig_box}, in this case the box covers almost the total time span of observations with a constant amplitude, with just a little decrease after BJD = 2460000. For the HD 176986 b and HD 176986 c, we see the signal to be stable throughout all the dataset.

With the current evidence, the drop in the apodized amplitude could be due to a combination of sampling, stellar activity, and the effectiveness of the YARARA correction. On the other hand, only further investigation can solve the puzzle of the origin of the variation we see in the apodized test applied to YARARA dataset.

.

\subsubsection{Single-instrument search for the 61.4 \si{\day} signal}

We conducted an informed search for the 61.4 \si{\day} signal in each of the three time series (H03,H15,HN) independently. This was done to check that the signal is not present only in one of the instruments. We considered for this test the activity-corrected dataset for simplicity, as the results between the datasets are consistent. Due to the limited nights of observations we have for each dataset, 124 for H03, 108 for H15, and 98 for HN, we considered for this test a model with a normal prior on the period of the planet.  We obtained, for the 61.4 \si{\day} signal, an amplitude K = 1.39 $\pm$ 0.34 \si{\meter\per\second} for H03, an amplitude K = 1.08 $^{+0.31}_{-0.32}$ \si{\meter\per\second} for H15, and an amplitude K = 1.51$^{+0.28}_{-0.26}$ \si{\meter\per\second} for HN. The H03 and H15 results are compatible at 0.67 $\sigma$, the H03 and HN results are compatible at 0.28 $\sigma$, and the H15 and HN results are compatible at 0.84 $\sigma$. The periods found for the outer signal are 61.42$^{+0.24}_{-0.21}$ \si{\day} in the H03 dataset, 61.15$^{+0.21}_{-0.18}$ \si{\day} in the H15 dataset, and 61.32$\pm$ 0.09 \si{\day} in the HN dataset. All the results are compatible within 1$\sigma$. 
We conclude that the signal is present in all the different datasets, and the different amplitudes we recover are related to the different sampling and activity phases of the star.
\subsection{Presence of detected signals in activity indicators}

Following the example of \citet{2025_stefanov_gj3998}, we searched for the planetary signals in the activity indicators. If the signal were of planetary origin, it would only appear in the RV time series. We ran two models with the same characteristics as the three-sinusoid model that we considered as our reference model, with the addition of a sinusoid with a uniform prior $\mathcal{U}$[40d,100d] and we were not able to find a significant signal at $\sim$ 61.4 \si{\day} in either the FWHM or the S index. For the FWHM we found a signal of amplitude K$_{FWHM}$ = 0.57$^{+0.59}_{-0.40}$ \si{\meter\per\second} and an undefined period of P$_{FWHM}$ = 63$^{+24}_{-18}$ \si{\day}.
For the S-index time series multiplied by 100, we found a signal of amplitude K$_{S-index}$ = 0.13$^{+0.11}_{-0.09}$ and a period of P$_{S-index}$ = 67$^{+25}_{-21}$ \si{\day}.
In both tests, we used the time series not corrected for activity, for the purpose of the test.

We repeated the same analysis for the other two planets at 6.49 \si{\day} and 16.8 \si{\day}, and we again found a similar result, with an amplitude of the sinusoid compatible with zero and an undefined period. For the 6.5 \si{\day} signal we considered a uniform prior on the period $\mathcal{U}$[5d,10d]. We found a signal  of amplitude K$_{FWHM}$ = 0.84$^{+0.55}_{-0.58}$ \si{\meter\per\second} and period P$_{FWHM}$ = 6.6$^{+2.0}_{-1.0}$ \si{\day} and K$_{S-index}$ = 0.14$^{+0.10}_{-0.09}$ and P$_{S-index}$ = 7.1$^{+1.3}_{-1.6}$ \si{\day} for FWHM and S$_{MW}$, respectively.
For the 16.8-\si{\day} signal we used a prior $\mathcal{U}$[10d,20d] and we found K$_{FWHM}$ = 0.49$^{+0.48}_{-0.35}$ \si{\meter\per\second} and P$_{FWHM}$ = 14.8$^{+3.2}_{-3.1}$ \si{\day} and K$_{S-index}$ = 0.077$^{+0.082}_{-0.054}$ and P$_{S-index}$ = 14.9$^{+3.3}_{-3.2}$ \si{\day}. 
We conclude that the planetary signals are not present in the activity indicators.
\section{Discussion}
\label{sec_dis}
\subsection{Planetary system}
\label{planetary_system_disc}
\begin{figure*}[!h]
    \begin{minipage}{0.9\textwidth}
        \includegraphics[width=\linewidth]{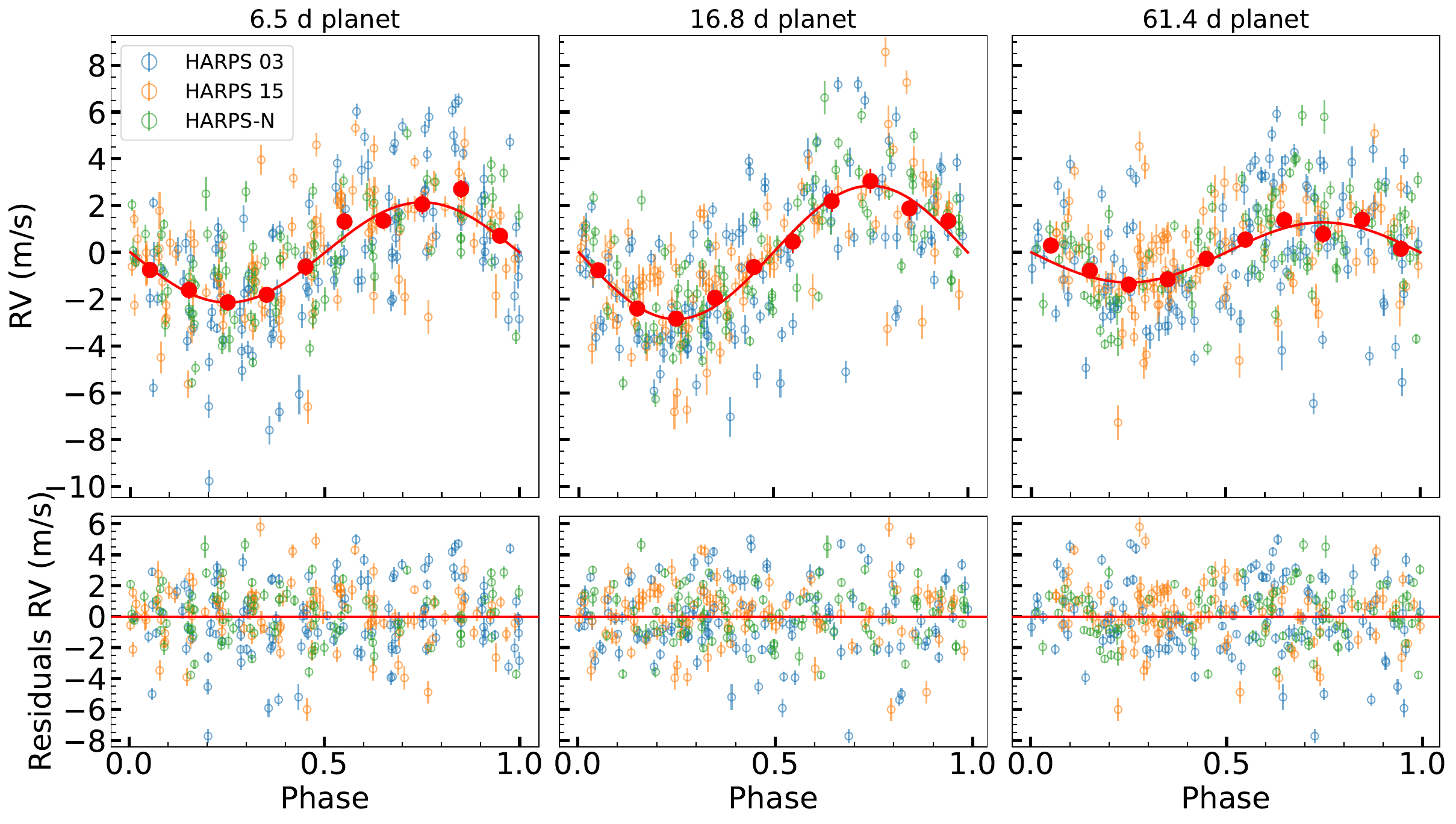}
    \end{minipage}
    \caption{Phase-folded plot of HD 176986 b, HD 176986 c, and HD 176986 d.   
    }
    \label{phase_folded_plot}
\end{figure*}

We revisited the planetary system of HD 176986, a K2.5 star intensively observed with HARPS and HARPS-N, which is known for hosting two super-Earths in a close orbit \citep{2018_ropes_alejandro}. We considered a larger dataset compared to the original work, growing from 234 to 330 nights of observations and with a longer baseline, from 13.2 years to 18.6 years. Furthermore, we extracted the RVs with YARARA, which corrects for systematics and stellar activity. We performed a FIP analysis to determine the number of signals of interest. We found three signals in the FIP periodogram exceeding the 1 \% FIP at periods of $\sim$ 6.5 \si{\day}, 16.8 \si{\day}, and 61.4 \si{\day}. 
The 61.4 \si{\day} period signal is a new planet that resides interior compared to the inner edge of the HZ. We compared the evidence of a two-planet and a three-planet model. We considered the RV time series alone with activity corrected by YARARA and multidimensional GP framework with RVs and activity indicators. The new planet is favored in terms of lnZ by a 
$\Delta$ lnZ = 16.1-17.7, depending on the dataset we consider. We discuss the difference between the datasets in Sect. \ref{yarara_act}. In this section, and when not explicitly stated, we are going to consider only the multidimensional GP analysis because this permits us to include the analysis of the stellar rotation period and magnetic cycle.
In our three-planet model we found K$_b$ = 2.14 $\pm$ 0.17 \si{\meter\per\second}, P$_b$ = 6.49164$^{+0.00030}_{-0.00029}$ \si{\day}, and the time of conjunction T0$_b$ = 2460624.06 $\pm$ 0.17 BJD. 
For planet c, we found K$_c$ = 2.84 $\pm$ 0.18 \si{\meter\per\second}, P$_c$ = 16.8124 $\pm$ 0.0015 \si{\day}, and time of conjunction T0$_c$ = 2460614.69 $_{-0.31}^{+0.32}$ BJD. For the new signal, which we refer to as HD 176986 d, we found K$_d$ = 1.28 $\pm$ 0.17 \si{\meter\per\second}, P$_d$ = 61.376$^{+0.051}_{-0.049}$ \si{\day}, and T0$_d$ = 2460640.5 $_{-3.1}^{+3.0}$ BJD.
We show in Fig. \ref{phase_folded_plot} the phase-folded plot of the three planets in the multidimensional GP model. 

HD 176986 d is a planet discussed for the first time in this work. It has a minimum mass of M$_d$$\sin$i$_d$ = 6.76$_{-0.92}^{+0.91}$ M$\oplus$. The minimum masses found for the two other planets are MM$_b$$\sin$i$_b$ = 5.36 $\pm$ 0.44 M$\oplus$ for HD 176986 b and M$_c$$\sin$i$_c$ = 9.75$_{-0.64}^{+0.65}$ M$\oplus$ for HD 176986 c. 
Masses of HD 176986 b and HD 176986 c are compatible within 1$\sigma$ with the masses found in \citet{2018_ropes_alejandro}. 
We resume in Table \ref{table_planets} the ephemeris of the three planets. We show in Table \ref{table_priors} the results of our best model and the priors we used.

\begin{figure}[!h]
    \begin{minipage}{0.45\textwidth}
        \includegraphics[width=\linewidth]{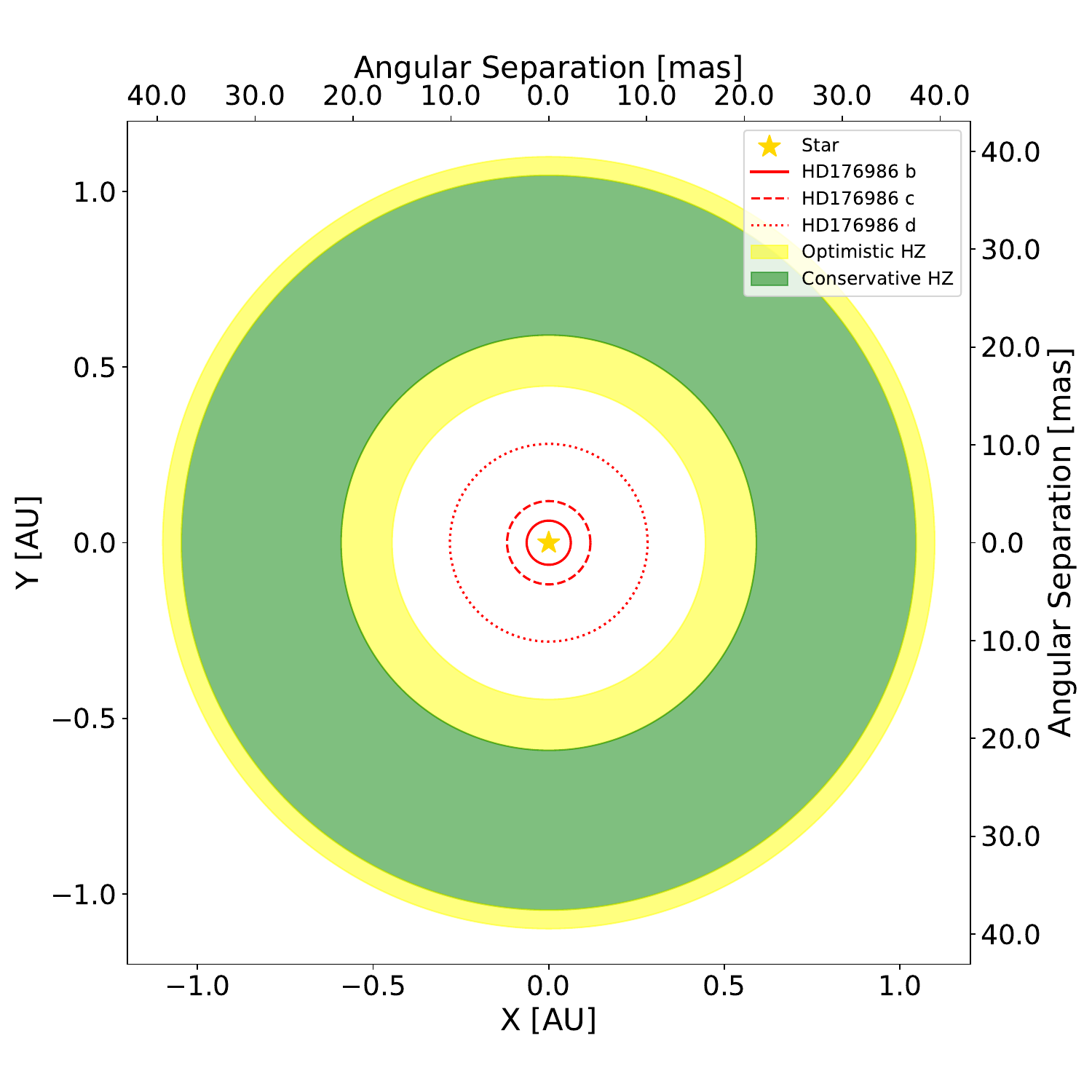}
    \end{minipage}
    \caption{Planetary system of HD 176986 relative to the HZ of the star.   
    }
    \label{hd176986_planetary_system}
\end{figure}

All the planets reside in an inner orbit compared to the HZ of the system. We plot in Fig. \ref{hd176986_planetary_system} the position of the planets of the system with respect to the HZ. HD 176986 b receives an insolation S$_b$ = 81.7 + $_{-8.6}^{+9.3}$ S$\oplus$, HD 176986 c receives an insolation S$_c$ = 22.9 + $_{-2.4}^{+2.6}$ S$\oplus$, and HD 176986 d receives an insolation S$_d$ = 4.09 $_{-0.43}^{+0.47}$ S$\oplus$. The equilibrium temperatures of the planets are T$_{eq b}$ = 767 $\pm$ 21 K, T$_{eq c}$ = 558 $\pm$ 15 K, T$_{eq d}$ = 363 $\pm$ 10 K. To calculate the temperature of the planets, we used Eq. \ref{temp_equation} \citep{2010_seager_temp}:
\begin{equation}
T_{\mathrm{eq}} = T_{\star}
\sqrt{\frac{R_{\star}}{2a}}
\left(1 - A\right)^{1/4}
\label{temp_equation}
.\end{equation}
We considered an albedo of 0.3.

HD 176986 d is an example of the capabilities of blind search RV surveys, as RoPES, to detect sub-Neptune planets with orbital periods of more than 50 \si{\day}. We compared our results with other detections of planets with minimum mass $<$ 20M$\oplus$ detected orbiting stars hotter than 4000K. We chose this cut to limit ourselves to stars similar to HD 176986. We compared the results obtained for HD 176986 with the literature results from the NASA Exoplanet archive \citep{2013_nasa_exoplanet_archive}. We limited the comparison to planetary systems hosting planets with insolation S comprising between 0.1 S$\oplus$ and 10 S$\oplus$ discovered by the RV method. Figure \ref{fig_comparison} shows the results. Only three planets reside inside the conservative HZ of their host star, with a total of 13 planets orbiting inside the optimistic HZ. The discovery of a planet like HD 176986 d, with an insolation of 4.09$^{+0.47}_{-0.43}$ S$\oplus$ and the shortage of sub-Neptunes in the HZ of hot stars points toward the need to continue long-term RV surveys to better explore the HZ of G-\&K-dwarfs.

The amplitude of the signal of HD 176986 b varies when considering the three instruments of our analysis independently (Sect. \ref{sec_apodized}). This creates the effect in the apodized test we see in Fig. \ref{fig_apodized}. Additional observations of the system are required to understand the origin of this variation, which could be related to sampling, stellar activity correction, although it is unlikely due to instrumental systematics being present in the dataset of all of our instruments. 

Due to the angular separation between the star and the planets, they are out of range for the future generation of
direct imaging facilities for atmospheric characterization.
A brief discussion on this topic is presented in Appendix \ref{sec_atmospheric_charact}.

\begin{figure}[htbp]
    \centering
    \begin{subfigure}[t]{0.48\textwidth}
        \centering
        \begin{tikzpicture}
            \node[anchor=north west, inner sep=0] (image) at (0,0) {\includegraphics[width=\linewidth]{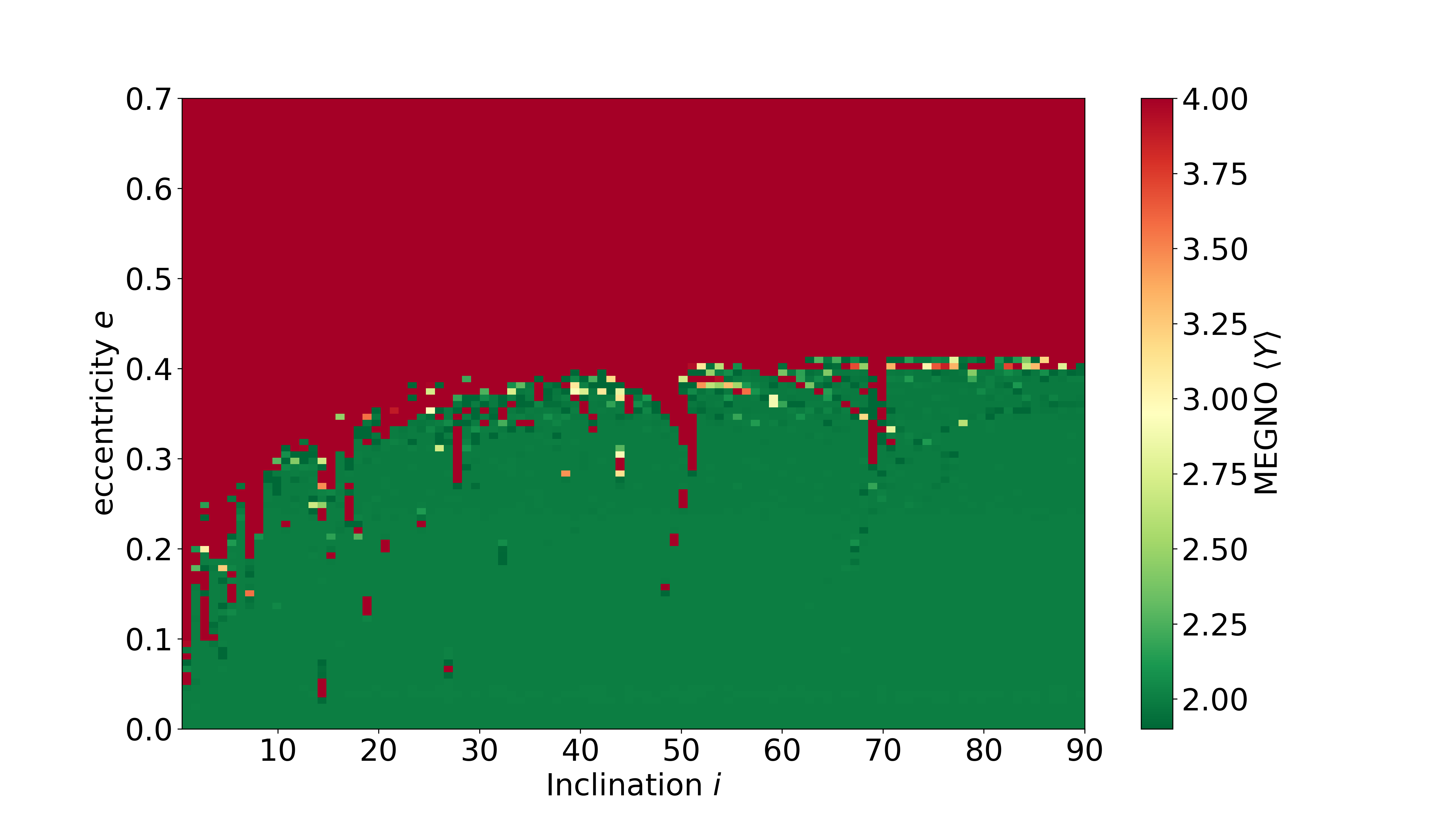}};
            \node[anchor=north west, inner sep=0] at (0.3, -0.5) {(a)}; 
        \end{tikzpicture}
        \label{fig:upper}
    \end{subfigure}
    \hfill
    \begin{subfigure}[t]{0.48\textwidth}
        \centering
        \begin{tikzpicture}
            \node[anchor=north west, inner sep=0] (image) at (0,0) {\includegraphics[width=\linewidth]{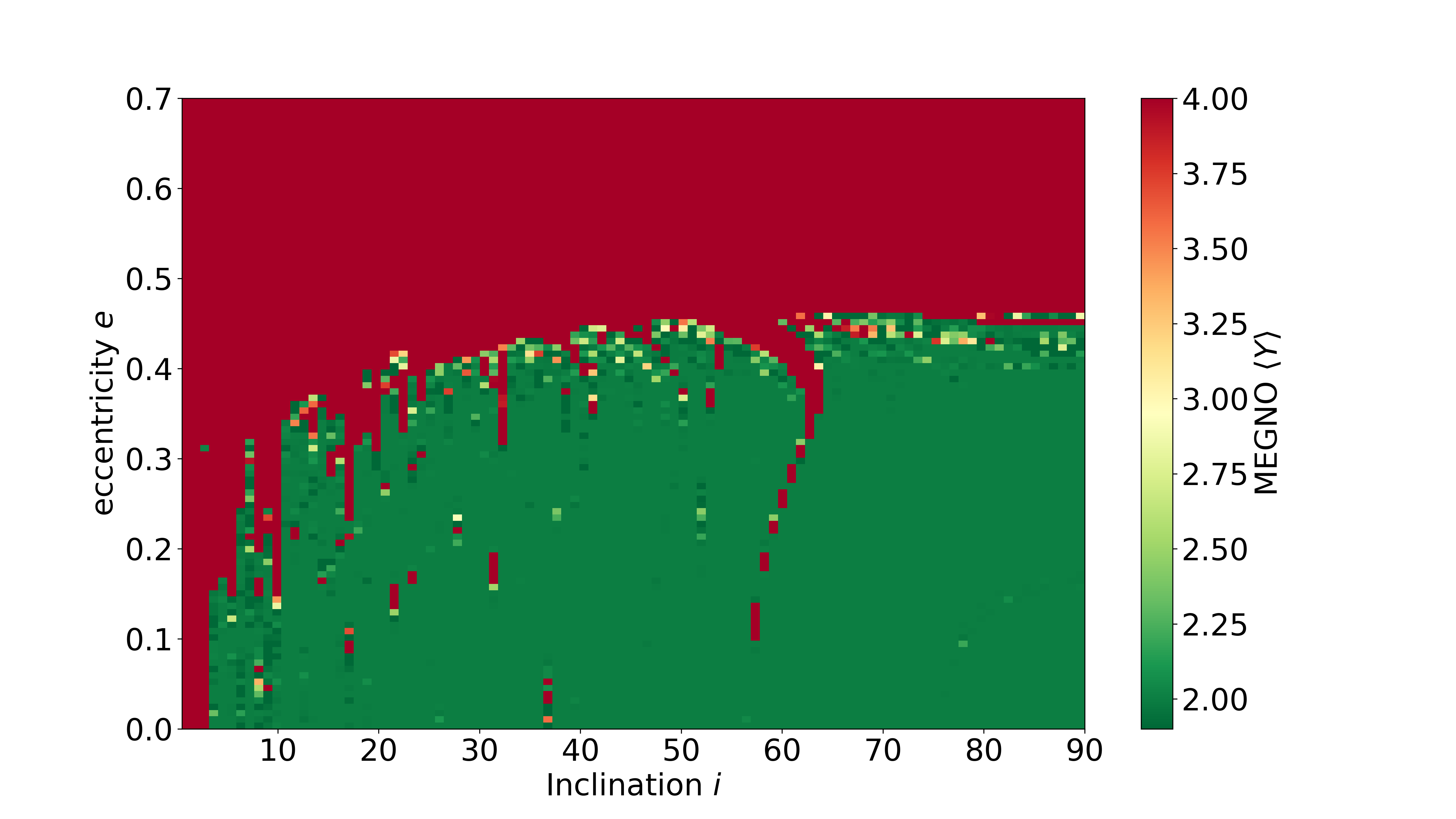}};
            
            \node[anchor=north west, inner sep=0] at (0.3, -0.5) {(b)}; 
        \end{tikzpicture}
        \label{fig:lower}
    \end{subfigure}
    \caption{Orbital stability tests for the planetary system of HD 176986. Panel (a): MEGNO stability map for the planetary system assuming a circular orbit for the two inner planets.
    Panel (b): MEGNO stability map for the planetary system assuming an eccentric (e=0.1) orbit for the two inner planets.}
    \label{fig_megno_circular}
\end{figure}

We tested the dynamical stability of the system with the code REBOUND \citep{2012_rein_rebound}. We built a stability map of the mean exponential growth of nearby orbits (MEGNO). A MEGNO value close to 2 indicates a stable orbit.  We calculated the MEGNO of the system for two different configurations. We integrated the system for 500 000 years to calculate the MEGNO.
First we considered a circular orbit for the two inner planets and calculated the MEGNO for a 100x100 grid of values of eccentricity of HD 176986 d and inclination (we assume the same value of the inclination for all the planets in the system). This permits us to calculate the MEGNO for the tentative true masses of all the planets and different configurations of eccentricity of the third planet. We show the map in Fig. \ref{fig_megno_circular} a). 
We repeated the test assuming an eccentric orbit of 0.1 for the two inner planets. This value is compatible with the upper limits on their eccentricity we found in our analysis. We show in Fig. \ref{fig_megno_circular} b) the results for the test with the eccentric model. We found the system to be stable with inclinations larger than 10 degrees. This is not conclusive on the true mass of the planets of the system. We found the system to be stable for eccentricity $<$ 0.4. This value is compatible with the results found for the case of circular orbits.

\subsection{Activity correction by YARARA}
\label{yarara_act}

We repeated all the tests of our analysis twice to compare the results of two different models: 1) A multidimensional GP applied to RV, FWHM, S index, with shared parameters of the MEP kernel as the timescale of evolution and the stellar rotation period (see Appendix \ref{stellar_activity_appendix} for a description of the MEP kernel). 2) An RV-only analysis in which we exploited the correction for the activity made by YARARA at the level of the spectra. 

\begin{figure}[htbp]
    \centering
    \begin{subfigure}[t]{0.48\textwidth}
        \centering
        \begin{tikzpicture}
            \node[anchor=north west, inner sep=0] (image) at (0,0) {\includegraphics[width=\linewidth]{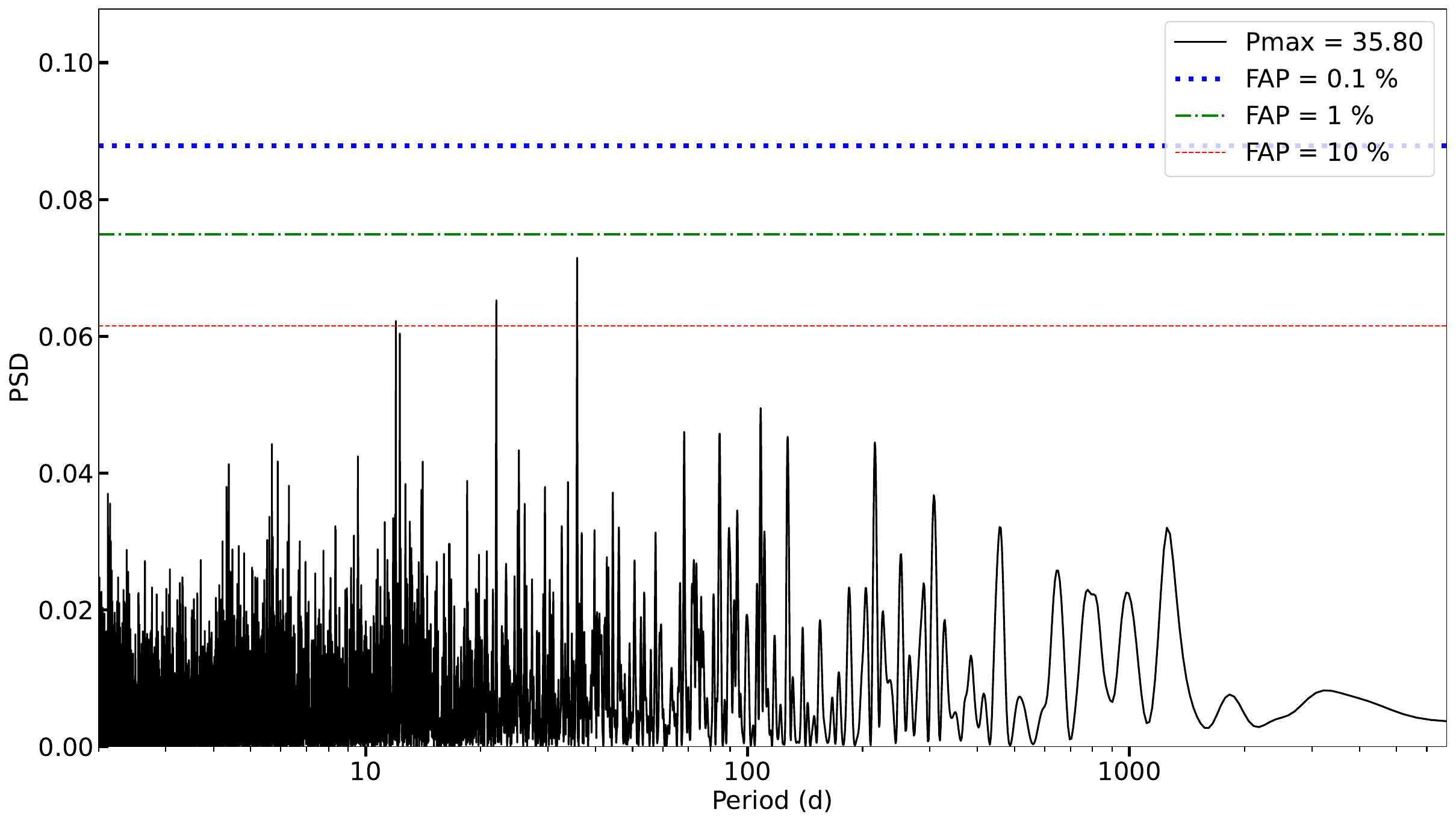}};
            
            \node[anchor=north west, inner sep=0] at (0.7, -0.4) {(a)}; 
        \end{tikzpicture}
        \label{fig:upper} 
    \end{subfigure}
    \hfill
    \begin{subfigure}[t]{0.48\textwidth}
        \centering
        \begin{tikzpicture}
            \node[anchor=north west, inner sep=0] (image) at (0,0) {\includegraphics[width=\linewidth]{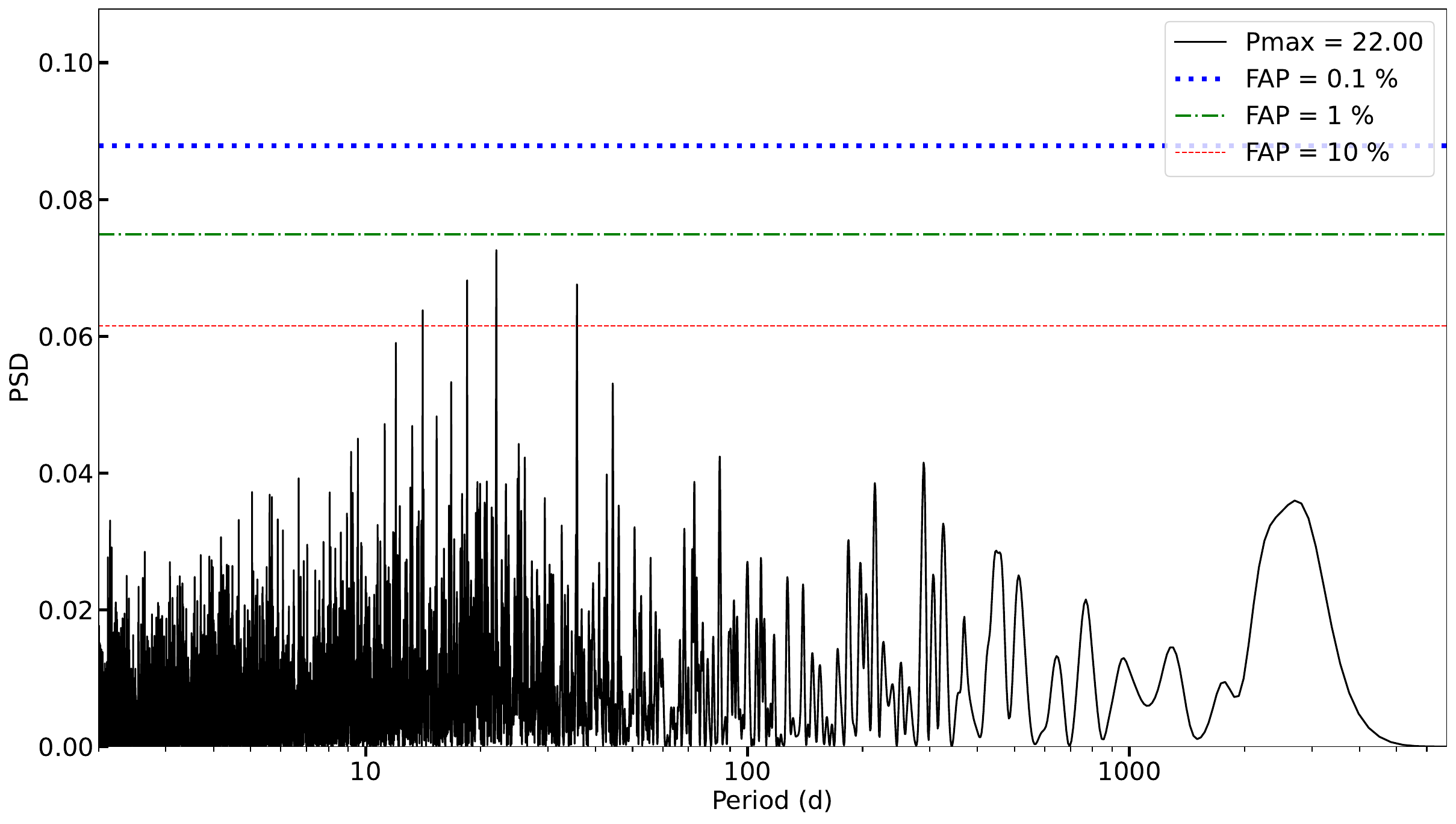}};
            \node[anchor=north west, inner sep=0] at (0.7, -0.4) {(b)}; 
        \end{tikzpicture}
        \label{fig:lower} 
    \end{subfigure}
    \caption{GLS periodogram of the residuals. Panel (a): GLS periodogram of the RV residuals after the subtraction of the planetary model along with the stellar activity model in the multidimensional GP framework.
    Panel (b): GLS periodogram of the RV residuals of the activity-corrected dataset after the subtraction of a three-planet model.}
    \label{residuals_periodogram}
\end{figure}

The two different analyses highlighted the model with three planets in a circular orbit as being the best model. The results on the amplitude of the three planets in the YARARA dataset corrected for activity are: K$_b$ = 2.03 $\pm$ 0.18 \si{\meter\per\second}, K$_c$ = 2.99 $\pm$ 0.17 \si{\meter\per\second}, and K$_d$ = 1.19 $_{-0.17}^{+0.16}$ \si{\meter\per\second}. All the results are 1$\sigma$ compatible with the results obtained for the multidimensional GP analysis. We note that the amplitudes for HD 176986 b and HD 176986 d are smaller than the one we found in the multidimensional GP analysis, while the amplitude for HD 176986 c is larger. We found a result compatible within 1$\sigma$ also for the period and phase of the signals. All the tests we conducted did not show a special preference between the datasets. We show in Fig. \ref{residuals_periodogram} a comparison between the GLS periodogram of the RV residuals of the two models. We see how the multidimensional GP model has some remnants in the GLS periodogram of the residuals at the rotation period of the star, while the strongest peak in the RV GLS periodogram of the residuals for the activity-corrected dataset is at $\sim$ 22 \si{\day}. Both the peaks have a FAP larger than 1 \% and in both cases, a model with an additional signal was disfavored in terms of evidence. 

In both cases, with and without the stellar activity correction, we found that the signal of HD 176986 becomes smaller in amplitude in the H15 dataset, as it was shown Sect. \ref{sec_apodized} on the apodized test. We do not find the same behavior in the analysis performed with the CCF-derived RVs. We cannot exclude the possibility that the signal is partially absorbed by the YARARA correction of systematics. 

\subsection{Need for multidimensional GP}
\label{multi_dim_sec}
We tested the need for multidimensional GPs. We subtracted the known planets from the data. Then, we injected different signals at the periods of planets b, c, and d in different runs, with amplitudes ranging from 0.2 \si{\meter\per\second} to 1.5 \si{\meter\per\second}. We fixed the period and phase of the injected planets and fit only for the amplitude. We left the nested sampler to determine parameters of the GP as the amplitude, the rotation period, the timescale of evolution, and the harmonic complexity. We imposed a uniform prior on the rotation period $\mathcal{U}$[25d,45d]. The prior on the timescale of evolution goes in logarithm space from half the lower limit of the prior on the rotation period, 12.5 \si{\day}, to roughly 1000 \si{\day}. We show in Fig. \ref{multi_dim_gp_overfitting} a comparison of the significance of the detection within the multidimensional GP framework and the 1D GP framework. For injected planets with the period of HD 176986 b, we observed that the two frameworks yield similar results in the detection, with a slightly better result for the 1D GP framework. For the period of HD 176986 c we observed the multidimensional GP framework to provide better results, and the same is more evident at the period of HD 176986 d. This shows the advantage of using a multidimensional GP framework to characterize long-period planets. An additional comparison was done, permitting the timescale of evolution to assume unphysical values, with a range in logarithm space between 0 \si{\day} and 7 \si{\day}. For this test, we considered a 1 \si{\meter\per\second} signal injected at the period of the newly discovered planet. For the 1D GP we found a value of the timescale of evolution of 2.57$_{-0.59}^{+0.70}$ \si{\day}, while for the multidimensional GP, we found 75$_{-17}^{+18}$ \si{\day}. Such a short value for the 1D GP is unrealistic for the timescale of evolution of the activity pattern, and it turns the GP non-periodic due to its flexibility. The short timescale of evolution of the 1D GP does not permit a proper definition of the rotation period, with Prot = 36.1$_{-7.2}^{+5.6}$ \si{\day}. In the multidimensional GP framework, we found Prot = 36.19$_{-0.71}^{+0.68}$ \si{\day}.

\subsection{Detection limits}
\label{det_lim}
We performed a sequence of recovery tests to explore the detection limits of the time series of HD 176986. We considered our best model for the multidimensional GP, both in RV and activity indicators. We subtracted from the RV time series the three planetary signals. We also considered the zero-points and jitters of the S index and FWHM datasets as fixed. We considered as free parameters of the model the zero-points and jitters for RV. We fit an additional sinusoidal, which is our test planet, to derive the sensitivity limit. We considered the parameters of the GP model as fixed from the previous analysis and applied them to the new time series of the residuals. The amplitude and the phase of the sinusoidal are free parameters of the model, while the period is taken from a grid of 1000 points covering the range between $\sim$ 1.6 \si{\day} and $\sim$ 5000 \si{\day} uniformly in logarithm. 

\begin{figure}[htbp]
    \centering
    \begin{subfigure}[t]{0.48\textwidth}
        \centering
        \begin{tikzpicture}
            \node[anchor=north west, inner sep=0] (image) at (0,0) {\includegraphics[width=\linewidth]{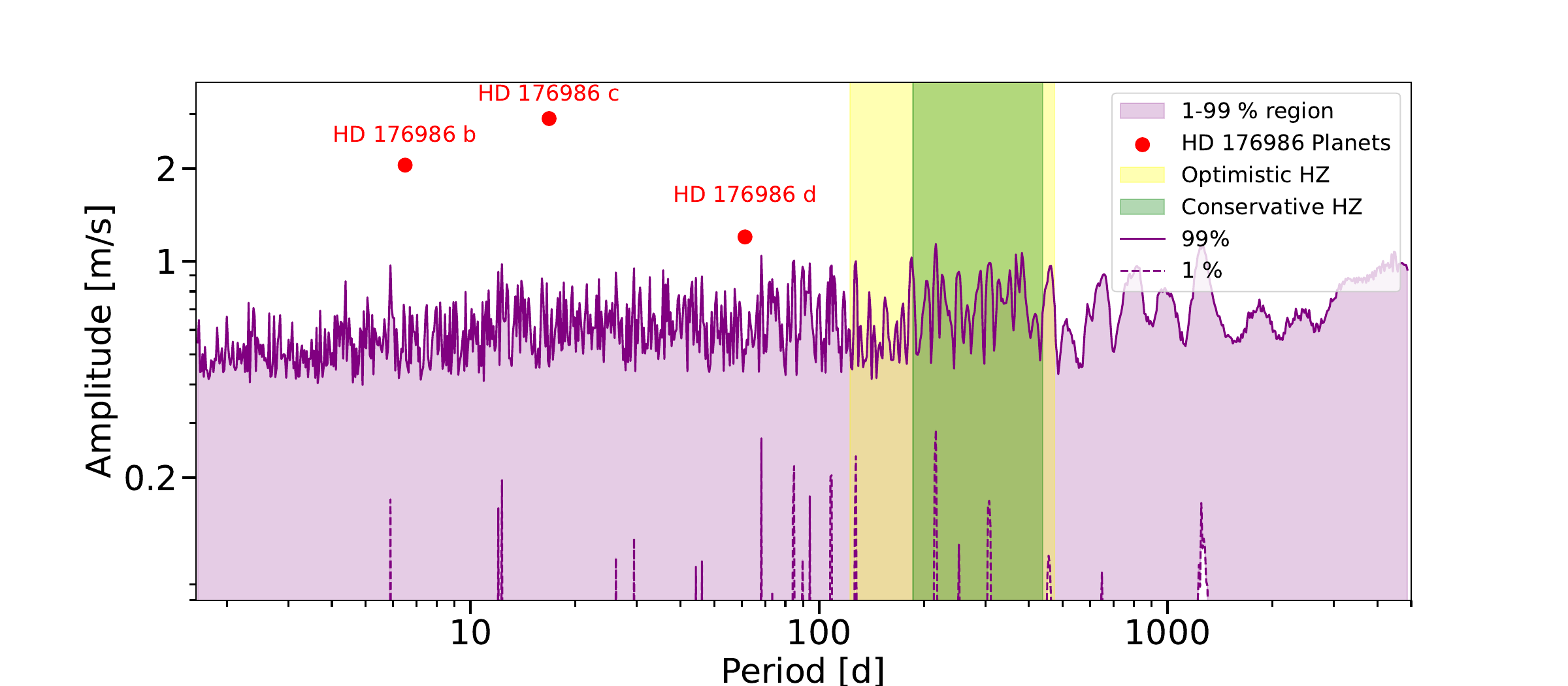}};
            \node[anchor=north west, inner sep=0] at (0.7, -0.4) {(a)}; 
        \end{tikzpicture}
        \label{fig:upper}
    \end{subfigure}
    \hfill
    \begin{subfigure}[t]{0.48\textwidth}
        \centering
        \begin{tikzpicture}
            \node[anchor=north west, inner sep=0] (image) at (0,0) {\includegraphics[width=\linewidth]{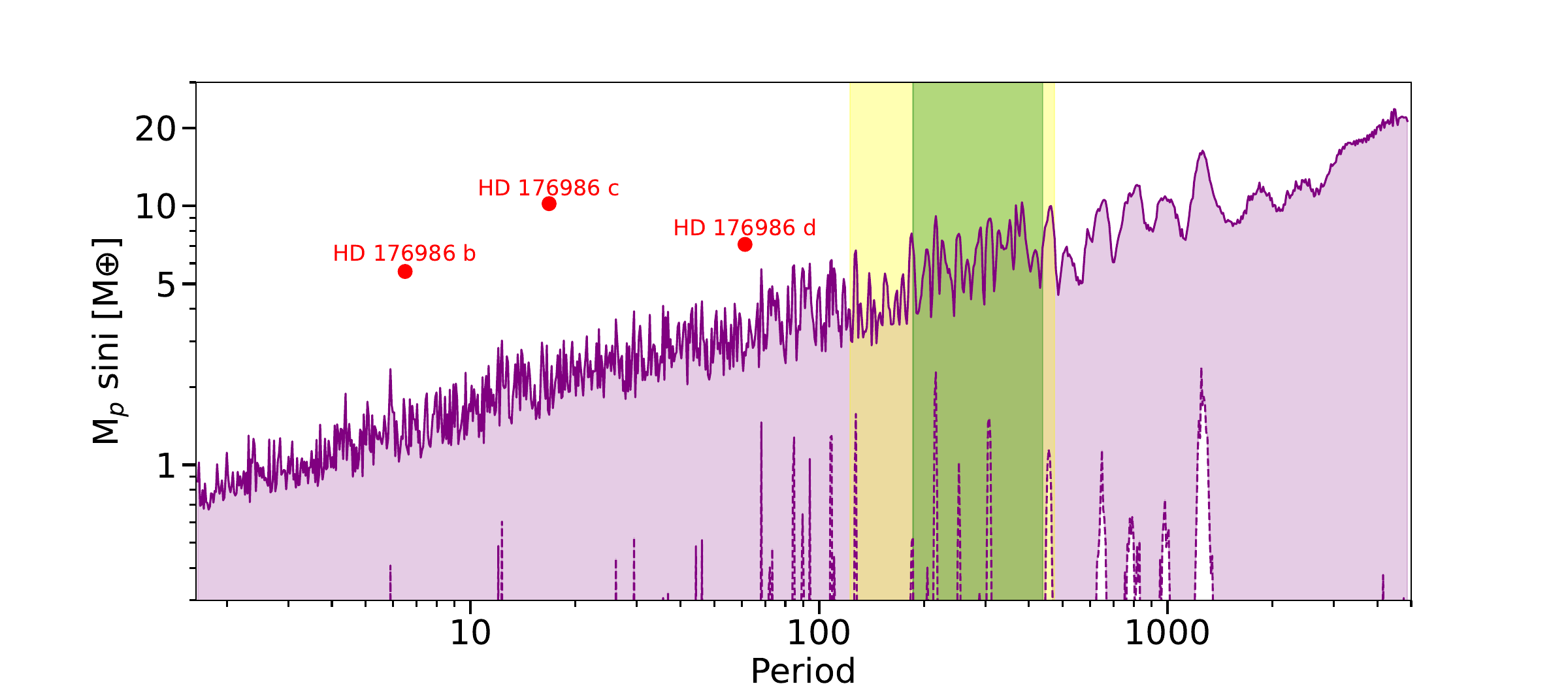}};
            \node[anchor=north west, inner sep=0] at (0.7, -0.4) {(b)}; 
        \end{tikzpicture}
        \label{fig:lower}
    \end{subfigure}
    \caption{Detection limits determination for HD 176986. Panel (a): Detection limits in amplitude for HD 176986. 
    Panel (b): Detection limits in mass for HD 176986.
    }
    \label{hd176986_detection_limits}
\end{figure}

The detection limit at every period is defined by considering the posterior of the additional signal we fit for. We considered the 99th percentile of the distribution in the amplitude as our detection limit at a specific period. A similar approach was followed in \citet{2014_tuomi_detection_limits} and \citet{2025_alejandro_proxima}. We show in Fig. \ref{hd176986_detection_limits} the detection limits in amplitude and mass. The region in light purple is the region comprising between 1\% and 99\% of the amplitude and mass. 

The mean amplitude we are sensitive to over the full dataset is 0.65 \si{\meter\per\second} considering the 99 \% upper limit. The median value is 0.63 \si{\meter\per\second}. The standard deviation is 0.16 \si{\meter\per\second}. 
We find a maximum for the amplitude at 99\% of 1.14 \si{\meter\per\second} and a maximum for the amplitude at 1 \% of 0.28 \si{\meter\per\second}. Both of them correspond to a $\sim$ 216 \si{\day} period on the grid. This period is of particular interest because it resides inside the HZ. We cannot claim a detection, and our FIP analysis did not point toward the presence of a significant power excess at that period. This period of interest can justify future follow-up of the system to investigate its nature.

We found a detection limit between 1 M$\oplus$ and 2 M $\oplus$ for orbital periods of less than 10 \si{\day}. Up to 100 \si{\day}, we are sensitive to planets with a minimum mass of more than 5 M$\oplus$. This limit increases up to 10 M$\oplus$ at 1000 \si{\day} orbital period and reaches a value comprising between 15 M$\oplus$ and 20 M$\oplus$ at longer orbital periods, between 2000 \si{\day} and 5000 \si{\day}. In the HZ of the star, we are sensitive to planets of minimum masses between 5 M$\oplus$ and 10 M$\oplus$. This result shows the capability of long-term RV surveys as RoPES, to explore the super-Earth regime in the HZ of G and K dwarfs. We can exclude the presence of Jupiter and Saturn-like planets orbiting around the star in orbits up to 5000 \si{\day}. For the full period range we are sensitive to sub-Neptunian planets in minimum mass. 
The detected planets are all within our detection capabilities.

\subsection{Stellar activity}
The joint analysis of the RVs with activity indicators gave us the possibility to derive the rotation period of HD 176986 and to determine the period of the magnetic cycle of the star. We found P$_{rot}$ = 36.05$^{+0.67}_{-0.71}$ \si{\day} and P$_{cycle}$ = 2432 $^{+64}_{-59}$ \si{\day}. The value we found for the rotation period in the multidimensional GP framework is 1$\sigma$ compatible with what we found in the analysis of S index and 1$\sigma$ compatible with the value reported in \citep{2018_ropes_alejandro}. The value for the cycle period in the multidimensional GP framework is compatible at 1.26 $\sigma$ with the value we found in the analysis of the S index alone. The star has a mean log$_{10}$(R'$_{hk}$) of $\sim$ -4.95. Log$_{10}$(R'$_{hk}$) has permitted us to estimate the rotation period of the star at P$_{rot}$ = 30.3$_{-5.7}^{+7.0}$ following the relationship $\log_{10}(P) = a \cdot \log_{10}(R'_{hk}) + b$ of \citep{2015_suarez_mascareno_rhk}. We used a = -0.773 $\pm$ 0.017 and b = -2.347 $\pm$ 0.002 for stars with solar metallicity. This value is 1$\sigma$ compatible with the value we found in our analysis.
We found a peak-to-peak amplitude related to the GP term in RV equal to 8.27 \si{\meter\per\second}. The derivative term is positively correlated with the RV term. Furthermore, the RV derivative term is larger in amplitude with a ratio of $\sim$ 2.9 with respect to the other term. This can point toward an activity which is dominated by the presence of dark spots instead of the inhibition of convective blueshift. 
We found a value for the timescale of evolution of the GP kernel $\lambda$ = 78$_{-18}^{+23}$ \si{\day}. This value is more than two times the rotation period of the star $\lambda$/Prot $\sim$ 2.17. This classifies HD 176986 as a “Beater” and not a “Sun-like” star following the classification of \citet{2017_giles_beater}. This could be related to the presence of polar spots, which are on average larger and last longer \citep{2017_giles_beater}.

\section{Conclusion}
We conducted a reanalysis of the RV datasets of the star HD 176986. HD 176986 is known from the literature to host two super-Earths at short orbital periods \citep{2018_ropes_alejandro}. In our analysis we took advantage of a larger dataset comprising all the observations we collected in the context of the RoPES program, spanning more than 6700 \si{\day} of observations. We extracted RVs with the LBL code YARARA, which corrects for instrumental systematics and, eventually, stellar activity. We retained 330 nights of RV observations split between HARPS and HARPS-N. We made a joint analysis of RVs with activity indicators in a multidimensional GP framework to mitigate the tendency of GP to overfit the data and characterize stellar activity. The joint analysis of activity indicators with RVs permitted us to determine the rotation period of the star, P$_{rot}$ = 36.05$^{+0.67}_{-0.71}$ \si{\day}, and the period of the stellar magnetic cycle, P$_{cycle}$ = 2432$^{+64}_{-59}$ \si{\day}. 

We recovered three planets orbiting the star. HD 176986 b and HD 176986 c are known in the literature, since \citet{2018_ropes_alejandro}. HD 176986 b has a semi-amplitude of K$_b$ = 2.14 $\pm$ 0.17 \si{\meter\per\second} and an orbital period of P$_b$ = 6.49164$_{-0.00029}^{+0.00030}$ \si{\day}. HD 176986 c has a semi-amplitude of K$_c$ = 2.84 $\pm$ 0.18 \si{\meter\per\second} and an orbital period of P$_c$ = 16.8124 $\pm$ 0.0015 \si{\day}. 

The third planet was discovered in our analysis. HD 176986 d has a signal amplitude of K$_d$ = 1.28 $\pm$ 0.17 \si{\meter\per\second} and an orbital period of P$_d$ = 61.376$_{-0.049}^{+0.051}$ \si{\day}. 
All the planets have a minimum mass of less than 10 M$\oplus$, with M$_b$ = 5.36 $\pm$ 0.44 M$\oplus$, M$_c$ = 9.75 $_{-0.64}^{+0.65}$ M$\oplus$, and M$_d$ = 6.76 $_{-0.92}^{+0.91}$ M$\oplus$.

We performed different tests, such as the apodized test, to assess the stability of the signal and research the signal at the period of the planetary signal in activity indicators, to verify the planetary origin of the signal associated with HD 176986 d. The tests suggested a planetary origin of the signal, and we confirm it as a new detected planet. 

We were able to perform a comparison between the activity correction in the multidimensional GP framework and the correction of the activity performed by YARARA. We recovered the same configuration for the planetary system, but we found the Keplerian model of the YARARA-corrected dataset to show a higher eccentricity, even if the model is not favored in $\Delta$ lnZ.

We tried to model the signals with Keplerians instead of sinusoidals but for all of the planets we were only able to put an upper limit on the eccentricity, with e$_b$ $<$ 0.09, e$_c$ $<$ 0.14, and e$_d$ $<$ 0.33. 
Furthermore, the models with a Keplerian solution were discarded in terms of the evidence of the model. A circular orbit model is the model favored by the evidence. 

We tested the sensitivity limit of our dataset. We found a mean sensitivity limit of 0.65 \si{\meter\per\second} throughout the dataset. Due to the lack of sensitivity, we cannot rule out the presence of 1 M$\oplus$ planets at any period. We are sensitive to planets with M $>$ 5 M$\oplus$ for orbital periods shorter than 100 \si{\day}. In the HZ of the star, we are sensitive to planets with masses between 5 M$\oplus$ and 10 M$\oplus$. The detection limit in mass rises up to 20 M$\oplus$ at orbital periods of thousands of days.

\label{sec_con}

\begin{acknowledgements}
      We thank the anonymous referee for the contribution given to improve the article. NN acknowledges funding from Light Bridges for the Doctoral Thesis "Habitable Earth-like planets with ESPRESSO and NIRPS", in cooperation with the Instituto de Astrofísica de Canarias, and the use of Indefeasible Computer Rights (ICR) being commissioned at the ASTRO POC project in the Island of Tenerife, Canary Islands (Spain). The ICR-ASTRONOMY used for his research was provided by Light Bridges in cooperation with Hewlett Packard Enterprise (HPE). 
      JIGH, AKS, RR, CAP, NN, VMP, and ASM acknowledge financial support from the Spanish Ministry of Science and Innovation (MICINN) project PID2020-117493GB-I00. The project that gave rise to these results received the support of a fellowship from the ”la Caixa” Foundation (ID 100010434). The fellowship code is LCF/BQ/DI23/11990071.
      This publication makes use of The Data \& Analysis Center for Exoplanets (DACE), which is a facility based at the University of Geneva (CH) dedicated to extrasolar planets data visualization, exchange, and analysis. DACE is a platform of the Swiss National Centre of Competence in Research (NCCR) PlanetS, federating the Swiss expertise in Exoplanet research. The DACE platform is available at \url{https://dace.unige.ch}.
      XD acknowledges the support from the European Research Council (ERC) under the European Union’s Horizon 2020 research and innovation programme (grant agreement SCORE No 851555) and from the Swiss National Science Foundation under the grant SPECTRE (No $200021\_215200$). This work has been carried out within the framework of the NCCR PlanetS supported by the Swiss National Science Foundation under grants $51NF40\_182901$ and $51NF40\_205606$.
      This research has extensively used the SIMBAD database operated at CDS, Strasbourg, France,
      and NASA’s Astrophysics Data System. This research has made use of NASA
      Exoplanet Archive, which is operated by the California Institute of Technology, under contract with the National Aeronautics and Space Administration
      under the Exoplanet Exploration Program. The manuscript was written using \texttt{Overleaf}. Extensive use of \texttt{numpy} \citep{2011_numpy} and \texttt{scipy} \citep{2020_virtanen_scipy}.
      The main analysis was performed in Python 3 \citep{python3_ref} running on a Ubuntu system \citep{ubuntu_2015}.
      Programs IDs for the HARPS observations we used in the analysis are: 072.C-0488(E), 091.C-0936(A), 183.C-0972(A), 192.C-0852(A), 198.C-0836(A), 106.21TJ.001, 108.22CE.001, 109.2392.001, 110.242T.00, 112.25SF.001, 113.26U2.001, and 114.27J9.001. Programs IDs for the HARPS-N observations we used in the analysis are: CAT14A\_83, CAT15A\_140, CAT16A\_109, CAT17A\_38, CAT18A\_115, CAT19A\_159, CAT20A\_121, and ITP15\_7. 
\end{acknowledgements}

%
%

\bibliographystyle{aa} %
\bibliography{bibliography_hd176986}


\begin{appendix}
\section{Methods}
\label{methods_sec}
For the parameter estimation, we used the nested-sampling tool Dynesty \citep{2020_destiny_speagle}. Dynesty gives an estimate of the natural logarithm of the evidence associated with a model, allowing for an easy model comparison. We considered as a criterion to accept a more complex model an improvement in lnZ of 5 or more. We used for the inference of parameters a number of livepoints equal to the minimum between 500 and 40 times the number of parameters of the model. For the FIP analysis we increased the number of livepoints to 100 times the number of parameters to better accomplish the degeneration of the model. We used it as a stopping criterion for convergence when the sampler remains with less than $\Delta$ lnZ = 0.01 to explore. To reckon the different zero-points of different instruments, we always consider an offset term for each instrument in the analysis. A jitter term was added in quadrature to the nominal error of the different instruments for each time series. The jitter term takes into account all the sources of noise we are not modeling for, and the instrumental noise. To remove outliers from different datasets, we binned the observations nightly and we apply a cut on the dataset consisting of a 3$\sigma$ clipping joined with the exclusion of measurements where the error is larger than three times the median error for each dataset. To implement Gaussian processes in our analysis we used S+LEAF \citep{2022_spleaf_delisle}. S+LEAF allows for only a certain kind of semi-separable matrix to be taken into account as a covariance matrix. In this way, the computational cost scales linearly with the dimension of the dataset, instead of with its cube, as it used to be in standard implementations. Before a fit for the jitter was available, exploratory GLS periodograms of our datasets were generated adding in quadrature a white noise term to the error on the measurements equal to the standard deviation of the dataset. We used the package for the GLS periodogram described in \citet{2009_gls_zeichmeister}.
\section{TESS analysis}
\label{tess_appendix}
\begin{figure}[!h]
    \begin{minipage}{0.45\textwidth}
        \includegraphics[width=\linewidth]{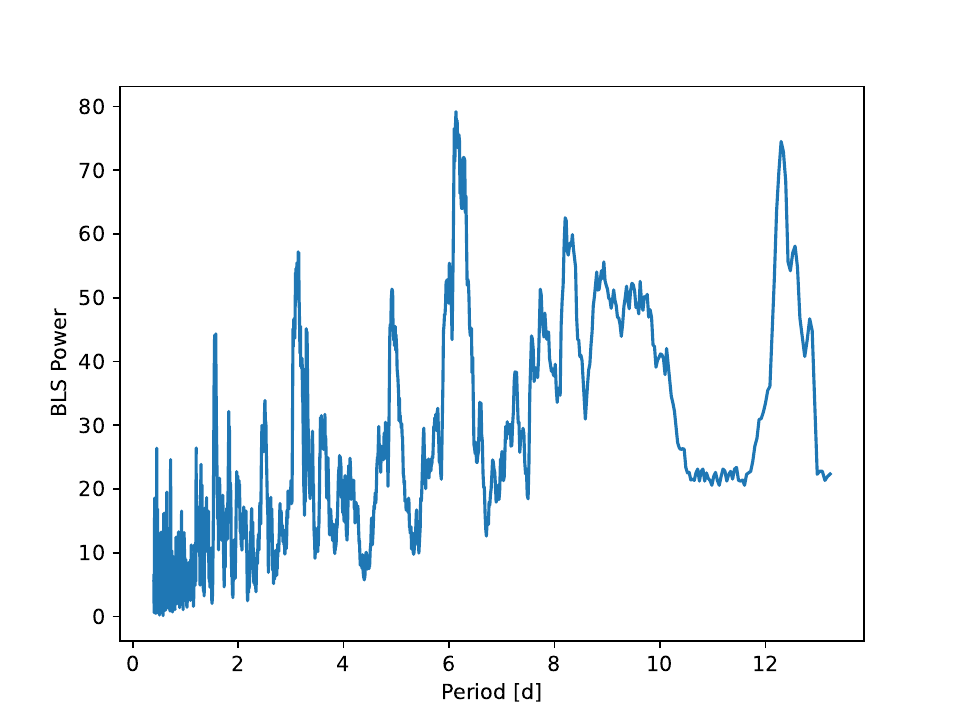}
    \end{minipage}
    \caption{BLS periodogram of TESS sector 80 data. There is no evidence of clear peaks in the periodogram related to a transit. There are no peaks related to the periods of planets found in RVs.}
    \label{TESS_BLS_periodogram}
\end{figure}

We used a BLS periodogram \citep{2002_kovacs_bls} to investigate the TESS time series searching for transit-like signals. We show in Fig. \ref{TESS_BLS_periodogram} the BLS periodogram for sector 80 of HD 176986. We did not find any significant periodicity in the BLS periodogram, nor hints for transit signals of the planets we found in RVs. 

\begin{figure}[!h]
    \begin{minipage}{0.45\textwidth}
        \includegraphics[width=\linewidth]{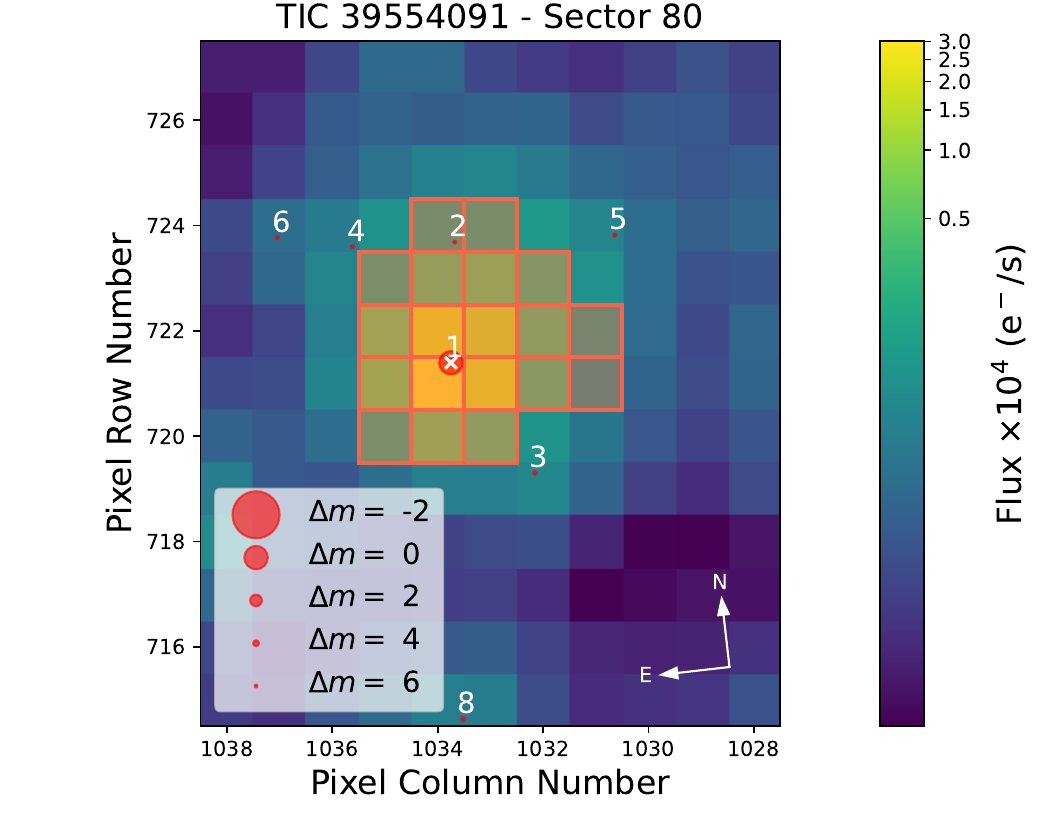}
    \end{minipage}
    \caption{TPF plotter. No other sources are inside the field of view of TESS within a $\Delta$ mag of 4. We can exclude contamination from the background star in the TESS light curves.}
    \label{TPF_plot}
\end{figure}

\section{ASAS-SN analysis}
\label{asassn_appendix}
\begin{figure}[htbp]
    \centering
    \begin{subfigure}[t]{0.48\textwidth}
        \centering
        \begin{tikzpicture}
            \node[anchor=north west, inner sep=0] (image) at (0,0) {\includegraphics[width=\linewidth]{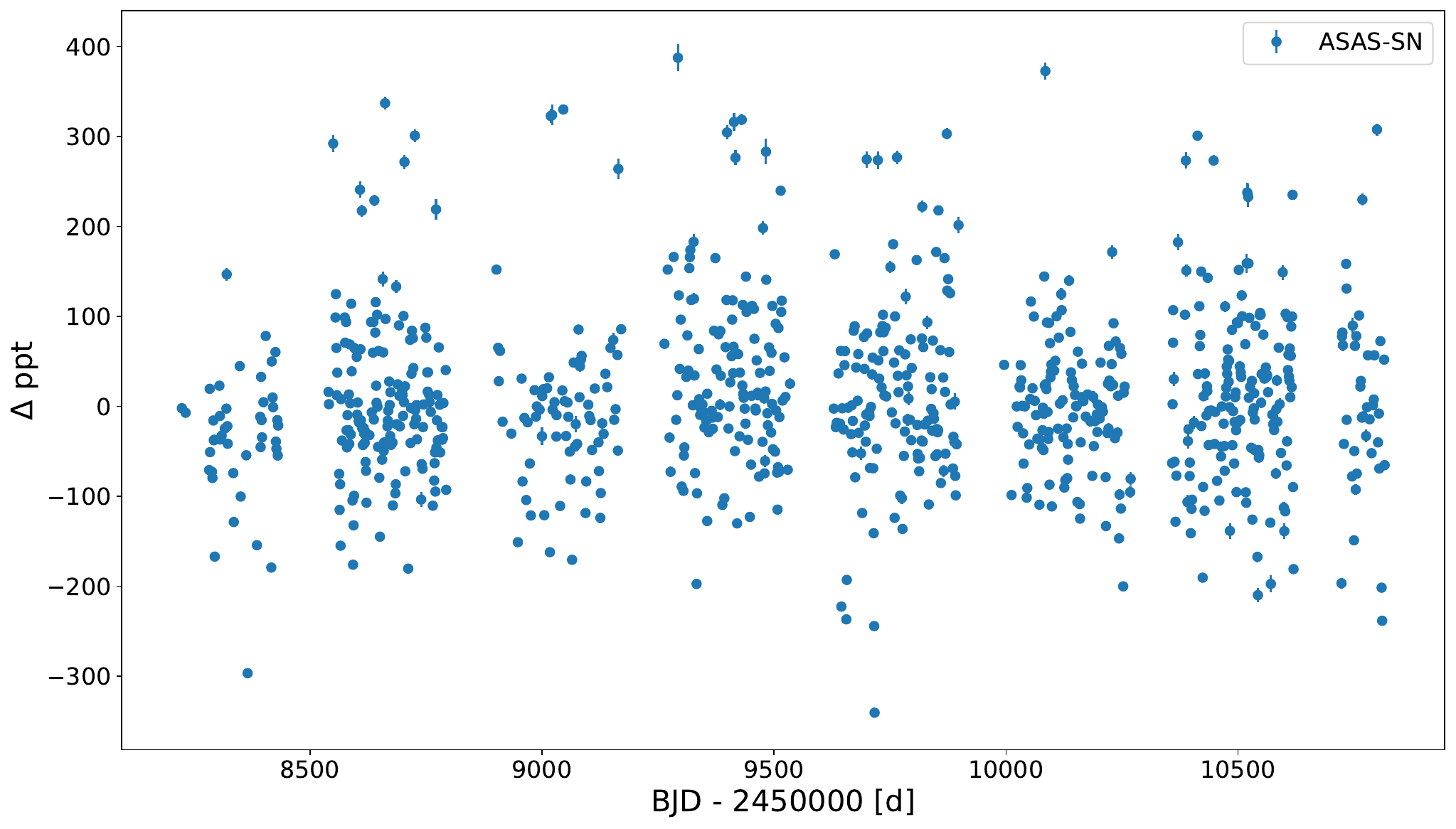}};
            \node[anchor=north west, inner sep=0] at (0.8, -0.4) {(a)}; 
        \end{tikzpicture}
        \label{fig:lower}
    \end{subfigure}
    \hfill
    \begin{subfigure}[t]{0.48\textwidth}
        \centering
        \begin{tikzpicture}
            \node[anchor=north west, inner sep=0] (image) at (0,0) {\includegraphics[width=\linewidth]{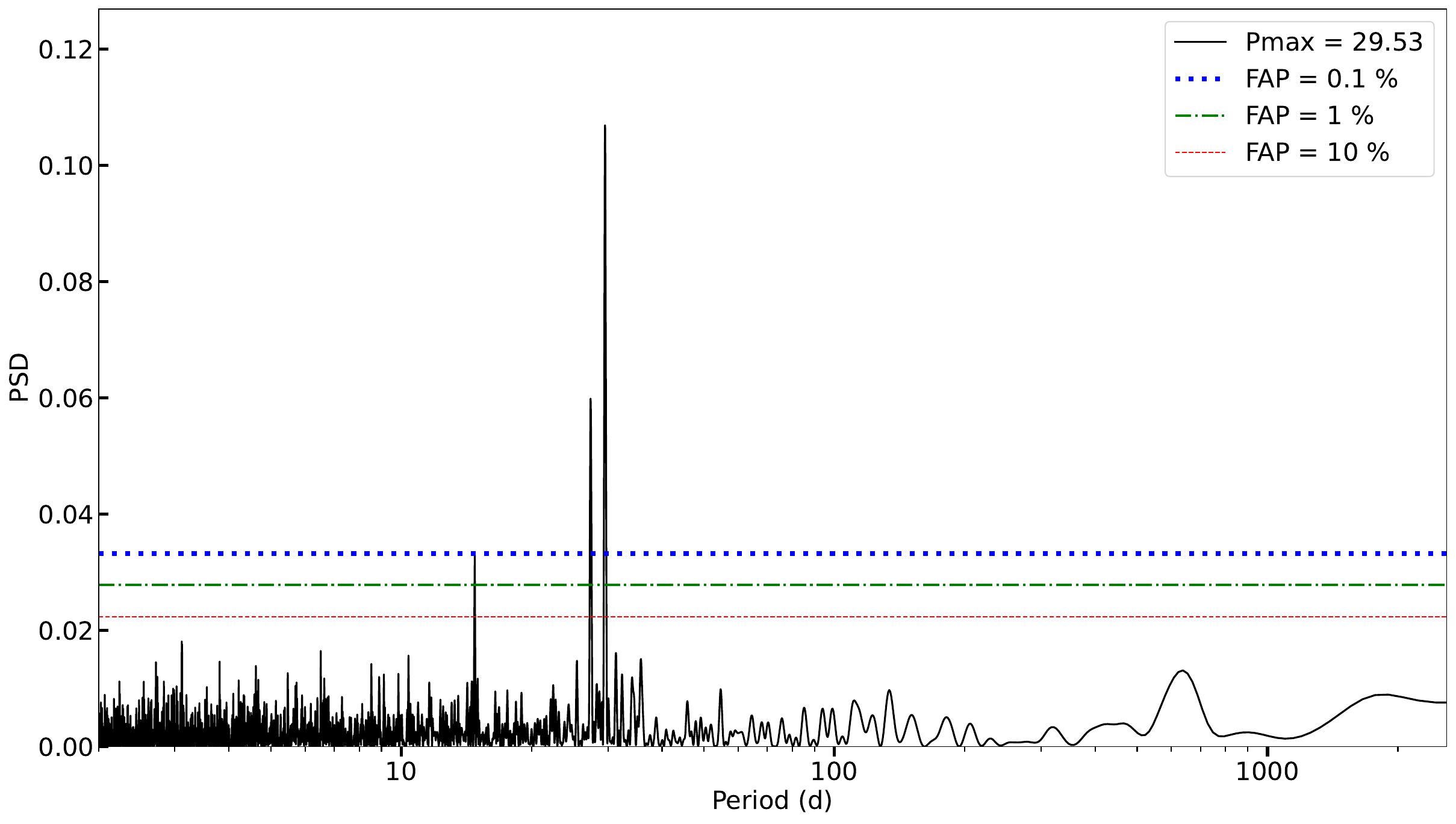}};
            \node[anchor=north west, inner sep=0] at (0.7, -0.4) {(b)}; 
        \end{tikzpicture}
        \label{fig:lower} 
    \end{subfigure}
    \caption{ASAS-SN dataset and GLS periodogram for HD176986. Panel (a): ASAS-SN time series for g-band filter.
    Panel (b): GLS periodogram of the ASAS-SN photometry collected in the g band. We see significant peaks at 29.56 \si{\day}, 14.78 \si{\day}, and 27.43 \si{\day}. These periods correspond with the synodic period of the Moon, half of the synodic period, and the sidereal period of the Moon.}
    \label{figure_asassn}
\end{figure}

We report here our analysis of the ASAS-SN photometric time series. We collected the data from the ASAS-SN Sky Patrol Photometry Database \citep{2014_asasn,2023_asassn_2}. We have photometry available from different cameras: bs, bk, bG, bo, bC, bH, and bp. We included in our analysis a total of 2948 single exposures in the g band on a total of 890 nights spanning 2592 \si{\day}. The standard deviation of the measurements is 93.9 ppt. In Fig. \ref{figure_asassn} we see the ASAS-SN time series and the relative GLS periodogram. 
We see strong peaks at the synodic period of the Moon, at its first harmonic, and at the sidereal period of the Moon. We modeled those with a sum of sinudoidal. Once we removed the three sinusouidal from our dataset, we can no longer see a peak in the GLS periodogram of the residuals with FAP $<$ 10 \%. 

\section{Stellar activity}
\label{stellar_activity_appendix}

Precise instruments at the level of 1 \si{\meter\per\second} in RV opened the way to the detection of Earth-like and super-Earth planets orbiting stars other than the Sun. When we reach this level of precision, effects from the star become detectable in the RV time series. Stellar systems are disequilibrium systems on different timescales. 

The oscillation of the external envelope of the star can generate short-term RV variations on the order of minutes for a Sun-like star with an amplitude of tens of \si{\centi\meter\per\second} \citep{2008_otoole_oscillations}.
To mitigate the effect of stellar pulsations we integrate our observations for a time longer than the timescale of the pulsations \citep{2011_dumusque_oscillations_granulations,2019_chaplin_oscillations}. 

Granulation phenomena arise from the convective flows happening in the outer layer of Sun-like stars. Granulation is responsible for RV variations on a timescale from a few minutes to multiple days with an amplitude up to a few meters per second \citep{2011_dumusque_oscillations_granulations,2011_mathur_granulation}. \citet{2011_dumusque_oscillations_granulations} proposes repeating the observation of a single target multiple times per night to average out the effect of granulation. Observability constraints make this method difficult to implement, especially in the case of large surveys. With HD 176986 this was not possible, so we could not implement a correction for the effect due to granulation. 

The Sun is known for showing an 11-year magnetic cycle \citep{1844_schwabe_solar_cycle}. This feature is observable in multiple stars other than the Sun \citep{2011_lovis_cycle}. Magnetic cycles can create signals up to tens of meters per second with a timescale of multiple years (cit). We modeled the cycle component in our time series with a sinusoidal and, if favored by the Bayesian evidence, additional harmonics.  

An important source of stellar activity to consider when searching for exoplanets in RV is the rotation of the star. The presence of spots on the surface of the star can create non-axisymmetric inhomogeneities in stellar flux and the magnetic fields associated with the spots can suppress the convection related to granulation \citep{1997_saar_activity}. The amplitude of the effect of stellar rotation on RV can reach tens of meters per second \citep{1997_saar_activity}. The RV variations induced by stellar rotation can produce a quasi-periodic pattern in the RV time series, mimicking or hiding Keplerian signals \citep{2021_meunier_activity}.  

To model the activity signal of stellar rotation we made use of Gaussian processes (GP) analysis \citep{2006_rasmussen_gp_book,2023_aigrain_gp}. Despite searching for a determined analytical function to interpolate the signal of activity, GPs are drawing samples from a distribution of functions that share a certain covariance function. We define the covariance function $\boldsymbol{K}$ = k($t_i$,$t_j$,$\phi$), where k($t_i$,$t_j$,$\phi$) is the covariance between observations made at time $t_i$ and $t_j$, and $\phi$ are the hyper-parameters of the covariance function. We define the covariance function as the kernel of the GP.

A common covariance function used in the field of exoplanets to model rotation-related activity is the quasi-periodic kernel defined as

\begin{equation}
k(t, t') = \sigma^2 \exp\left(-\frac{(t - t')^2}{2\lambda^2}\right) \exp\left(-\frac{\sin^2\left(\frac{\pi |t - t'|}{P}\right)}{2\Gamma^2}\right)
,\end{equation}where $\sigma$ is the amplitude of the kernel, $\lambda$ is the timescale of evolution of the correlations, P is the rotation period, and $\Gamma$ is the harmonic complexity. We used in our analysis an approximation of this kernel, the MEP kernel. The MEP kernel has the advantage of being representable as an S+LEAF matrix, so a matrix that can be represented as a sum of semi-separable and a LEAF matrix \citep{2020_spleaf_delisle}.  This decreases the computational cost of the likelihood in the process going from $\mathcal{O}(n^3)$ to $\mathcal{O}(n)$.

The GP method tends to overfit the dataset it is trained on, potentially absorbing an eventual planetary signal together with the activity. To mitigate this effect it is possible to model the RV time series together with one or more additional time series of activity indicators. The addition of activity indicators time series helps to constrain the parameter of the GP to be less prone to overfit the planetary signals. The multidimensional GP method is presented in \citet{2015_rajpaul_multigp} and an example of its effectiveness is given in \citet{2023_barragan_multigp}. The multidimensional GP framework was inspired by the FF' method \citep{2012_aigrain_ff'}. The FF' method models the RV variation with a linear combination of the squared flux obtained in simultaneous photometry and the flux multiplied by its time derivative: 

\begin{equation}
\text{RV} = V_r F(t) \dot{F}(t) + V_c F^2(t)
\end{equation}

The term with the derivative represents the flux modulation coming from the break of the flux symmetry and the term with $F^2(t)$ is linked to the suppression of convective blueshift in magnetized regions.

In the multidimensional GP framework, both RV and activity indicators are modeled with a linear combination of an underlying GP and its first derivative:

\begin{equation}
\begin{aligned}
\text{RV} &= V_c G(t) + V_r \dot{G}(t), \\
\text{A$_1$} &= A_{1c} G(t), \\
\text{A$_2$} &= A_{2c}G(t) + A_{2r} \dot{G}(t),
\end{aligned}
\end{equation}where V$_c$, V$_r$, A$_{1c}$, A$_{2c}$, and A$_{2r}$ are the coefficient of RV and two hypothetical activity indicators. The presence of one indicator without a term for the derivative and a second indicator with a term for the derivative is just a matter of representation: the number of activity indicators and the need for a derivative in the model are problem-dependent. 
We used different activity proxies in our analysis to characterize the stellar activity. We analyzed the time series of FWHM (DLW), Mount Wilson S index, H-$\alpha$ line activity, bisector span, and contrast.

\subsection{Full width at half maximum}
\begin{figure}[htbp]
    \centering
    \begin{subfigure}[t]{0.48\textwidth}
        \centering
        \begin{tikzpicture}
            \node[anchor=north west, inner sep=0] (image) at (0,0) {\includegraphics[width=\linewidth]{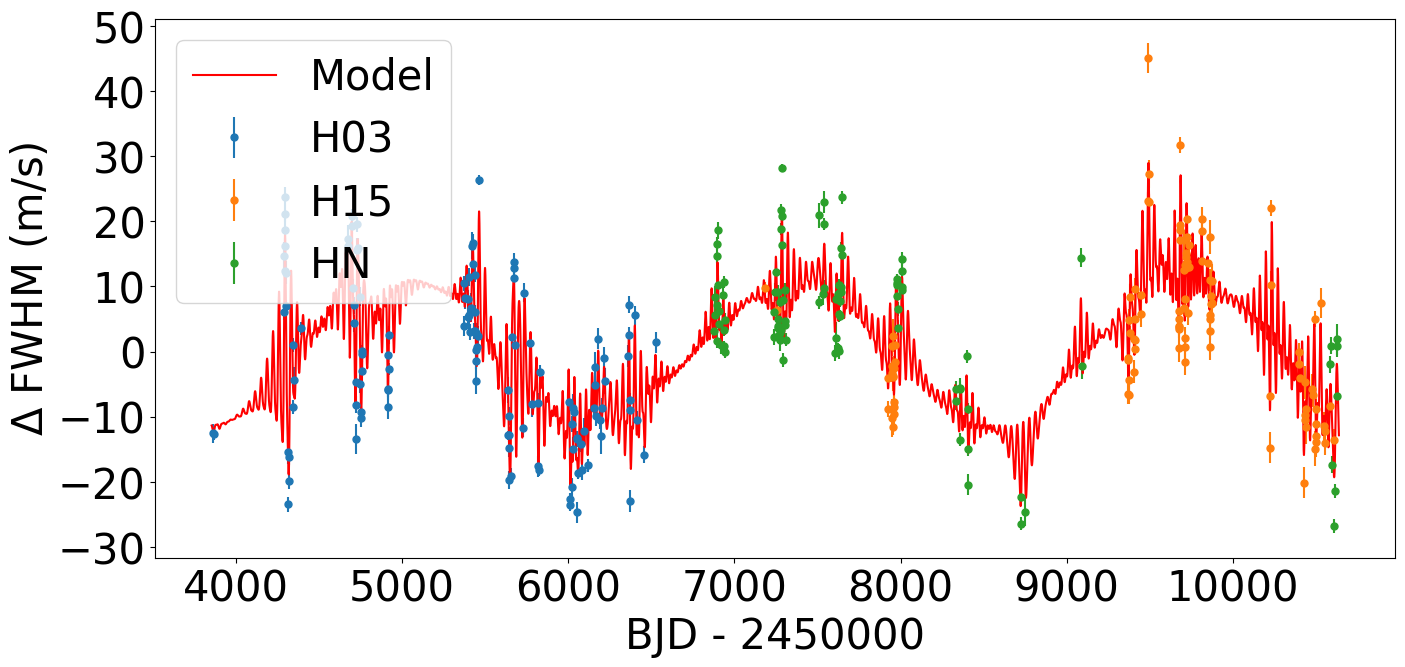}};
            \node[anchor=north west, inner sep=0] at (0.0, -0.4) {(a)}; 
        \end{tikzpicture}
        \label{fig:upper} 
    \end{subfigure}
    \hfill
    \begin{subfigure}[t]{0.48\textwidth}
        \centering
        \begin{tikzpicture}
            \node[anchor=north west, inner sep=0] (image) at (0,0) {\includegraphics[width=\linewidth]{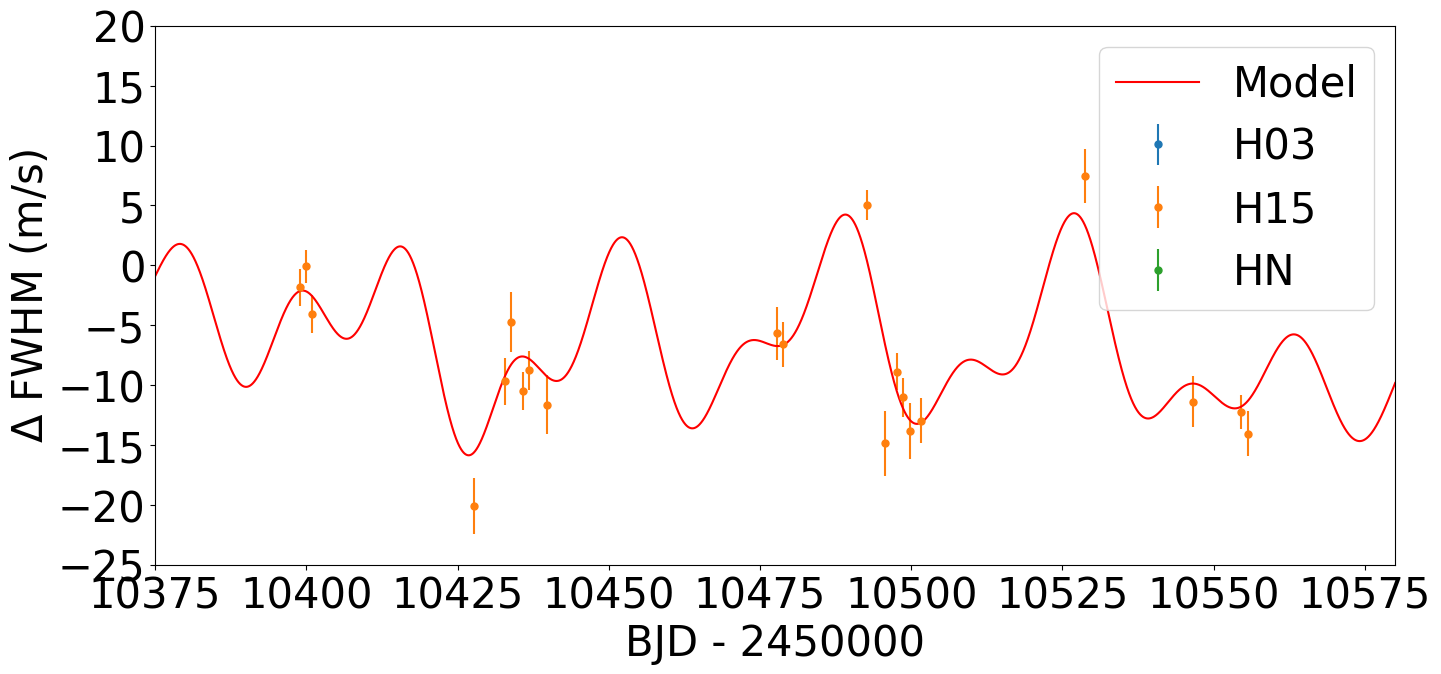}};
            \node[anchor=north west, inner sep=0] at (0.0,-0.4) {(b)}; 
        \end{tikzpicture}
        \label{fig:lower} 
    \end{subfigure}
    \caption{FWHM model for HD 176986. Panel (a): Global model of FWHM. It is possible to see the long-term modulation due to the magnetic cycle and the short-term variation related to stellar activity.
    Panel (b): Zoom of the activity model for FWHM. In the figure, one can see the activity pattern related to stellar activity.}
    \label{fig_fwhm_stellar_activity}
\end{figure}

The FWHM of the CCF is directly provided by the DRS and it has been proven to be an effective proxy for stellar activity in multidimensional GP regression \citep{2023_suarez_gj1002,2024_barnard_jonay}.
In Fig. \ref{fig_fwhm_stellar_activity} we show the activity model for FWHM as an example of the combination of a model for long-term magnetic cycles and stellar rotation. 

\subsection{S index}

S index or S$_{MW}$ is an indicator related to the intensity of the chromospheric emission in the Ca II H \& K lines. We used the same method of \citet{2011_lovis_cycle} to derive the S index. We defined two triangular passbands centered around the center of the H emission line at 3968.469 $\AA$ and the center of the K emission line at 3933.663 $\AA$. Both passbands have an FWHM of 1.09 $\AA$. We defined two rectangular passbands in the continuum as a reference for the emission line. One of them is centered at V = 3901.070 $\AA$ and the other is centered at R = 4001.070 $\AA$. Both have an FWHM of 20 $\AA$. The equation for the S index is
\begin{equation}
S_{\text{MW}} = \alpha \cdot \frac{\tilde{N}_H + \tilde{N}_K}{\tilde{N}_V + \tilde{N}_R} 
,\end{equation}where $\tilde{N}_H$, $\tilde{N}_K$, $\tilde{N}_V$, and $\tilde{N}_R$ are, respectively, the mean fluxes in the H, K, V, and R passbands. $\alpha$ and $\beta$ are two constants derived from the calibration of the method and are $\alpha$ = 1.111 and $\beta$ = 0.153. For a detailed review of the effects of spots and plages on the shape of Ca II H \& K lines, we refer to \citet{2024_cretignier_calcium}.
This indicator has proven to be effective in the analysis of stellar activity to retrieve stellar cycles and the rotation period of the star (as an example see \citealt{2019_mittag_sindex}). \citet{2015_suarez_mascareno_rhk} demonstrated that Log R'$_{HK}$, which is strongly related to the S index, is a proxy of the stellar rotation period.

\subsection{H$\alpha$}

The H$\alpha$ index can be a good proxy of stellar activity. It measures the strength of the H$\alpha$ emission line, which can also be an absorption line for weakly active stars. This activity indicator has been used intensively in the past for M-dwarfs \citep{2017_suarez_mascareno_halpha,2022_schofer_ha} but it can also be used as an activity proxy for G \& K stars. We calculated the H$\alpha$ index comparing the flux in the core of the H$\alpha$ line and the continuum, as explained in \citet{2011_gomes_da_silva_halpha}. 

We calculated the average flux over a rectangular passband with an FWHM of 1.6 $\AA$ centered at the H$\alpha_{core}$ = 6562.808 $\AA$. We divided this value for the averaged flux of two rectangular passbands: one centered at 6550.87 $\AA$  with an FWHM of 10.75 (H$\alpha_{V}$) and the other centered at 6580.31 $\AA$ with an FWHM of 8.75 $\AA$ (H$\alpha_{R}$). We used the equation 

\begin{equation}
H\alpha = \frac{H\alpha_{core}}{H\alpha_{V} + H\alpha_{R}}
\end{equation}

Where H$\alpha_{core}$, H$\alpha_{V}$, and H$\alpha_{R}$ are the averaged fluxes in the respective passbands.

\subsection{Bisector}

We consider as an activity indicator the asymmetry of the line-profile of the CCF. This indicator has proven to be an effective proxy for the stellar activity since the beginning of the field \citep{2001_queloz_bisector}.
\subsection{Contrast}
The contrast of the CCF is given by the relative ratio between the center of the CCF and the wings. This value can change due to the presence of stellar activity.

\newpage
\section{Box test}

\begin{figure}[htbp]
    \begin{subfigure}[t]{0.48\textwidth}
        \centering
        \begin{tikzpicture}
            \node[anchor=north west, inner sep=0] (image) at (0,0) {\includegraphics[width=\linewidth]{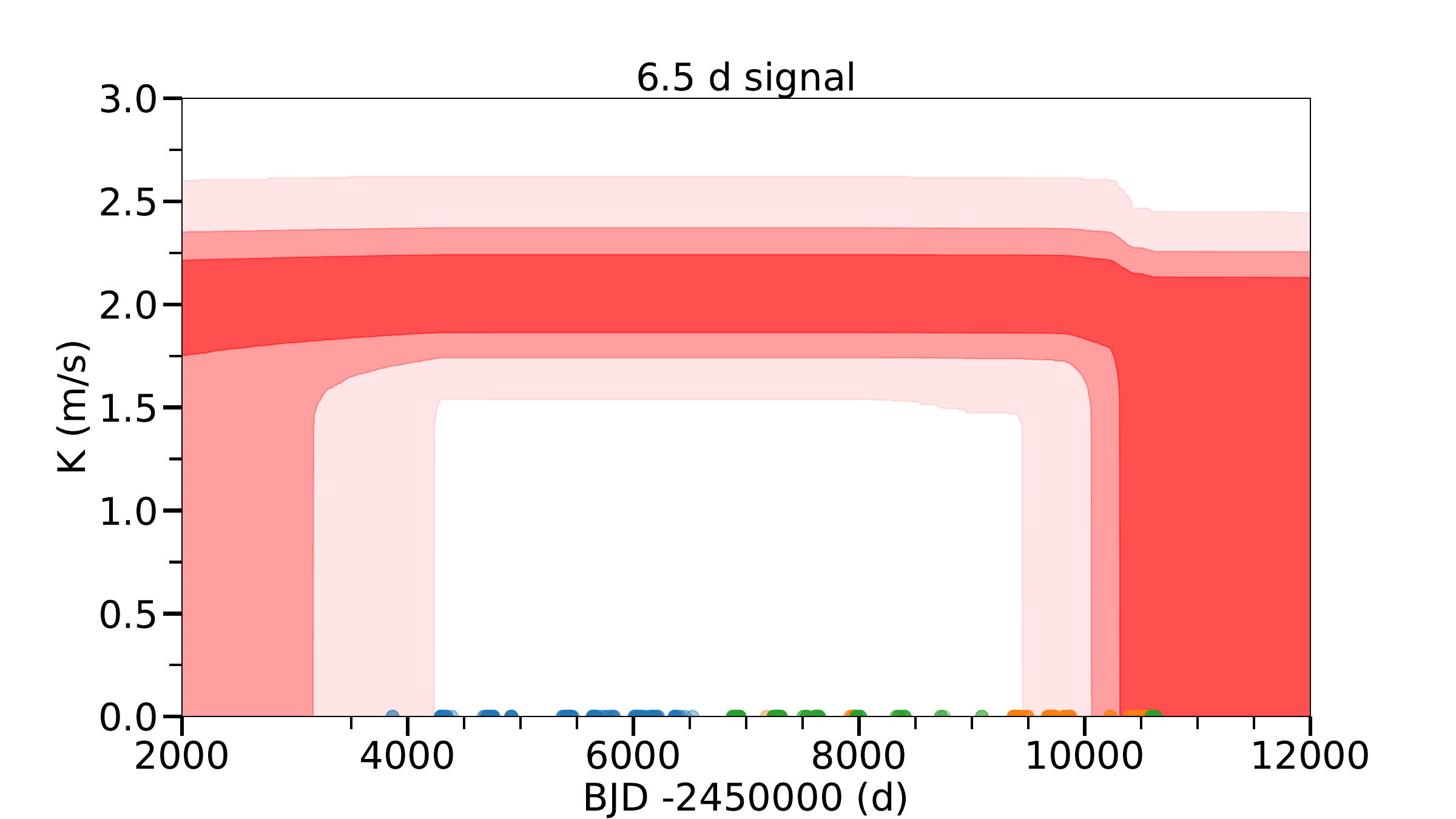}};
            \node[anchor=north west, inner sep=0] at (0.3, -0.6) {(a)};
        \end{tikzpicture}
        \label{fig:lower} 
    \end{subfigure}
    \hfill
    \begin{subfigure}[t]{0.48\textwidth}
        \centering
        \begin{tikzpicture}
            \node[anchor=north west, inner sep=0] (image) at (0,0) {\includegraphics[width=\linewidth]{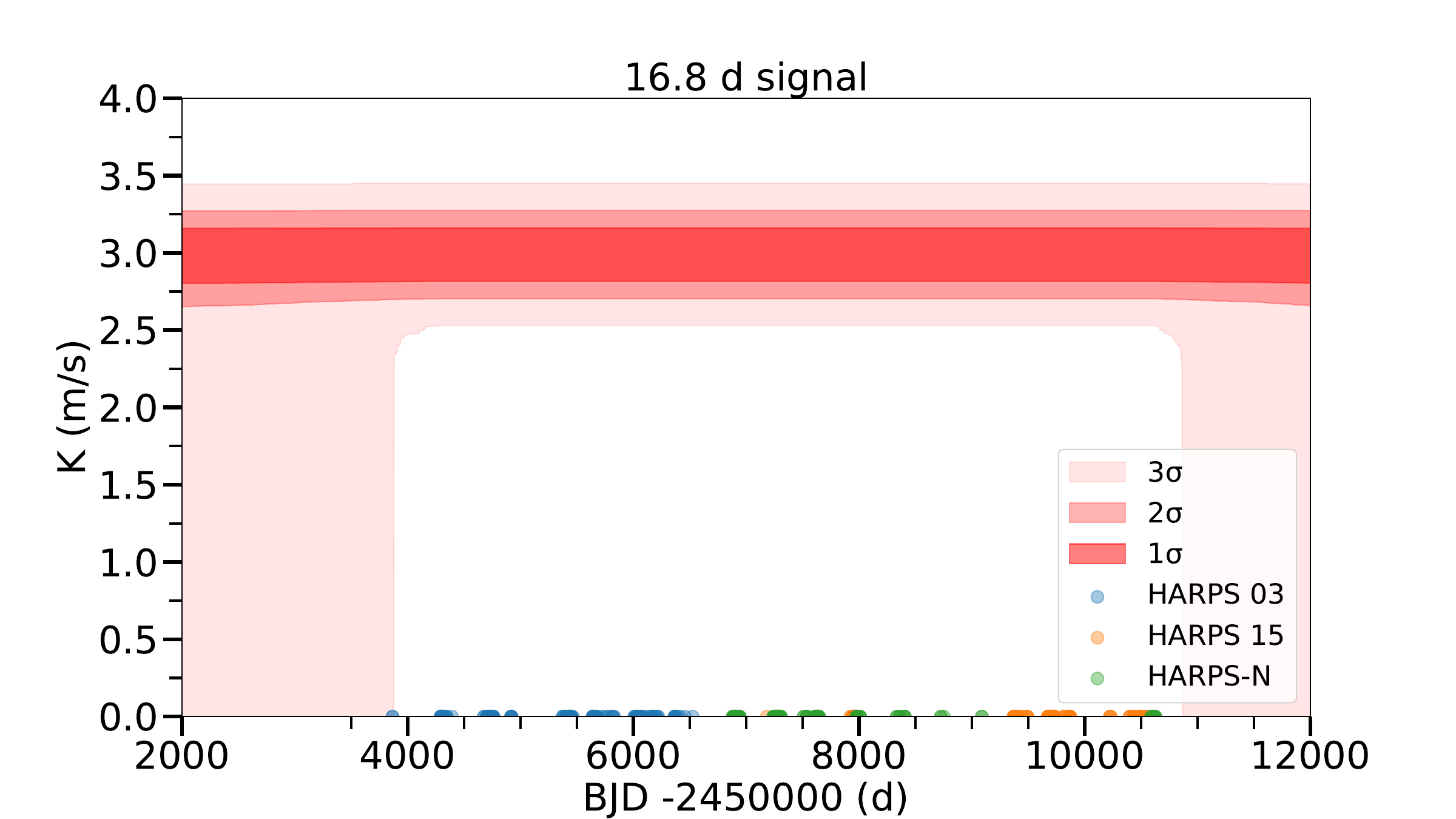}};
            
            \node[anchor=north west, inner sep=0] at (0.3, -0.6) {(b)}; 
        \end{tikzpicture}
        \label{fig:lower}
    \end{subfigure}
    \hfill
    \centering
    \begin{subfigure}[t]{0.48\textwidth}
        \centering
        \begin{tikzpicture}
            \node[anchor=north west, inner sep=0] (image) at (0,0) {\includegraphics[width=\linewidth]{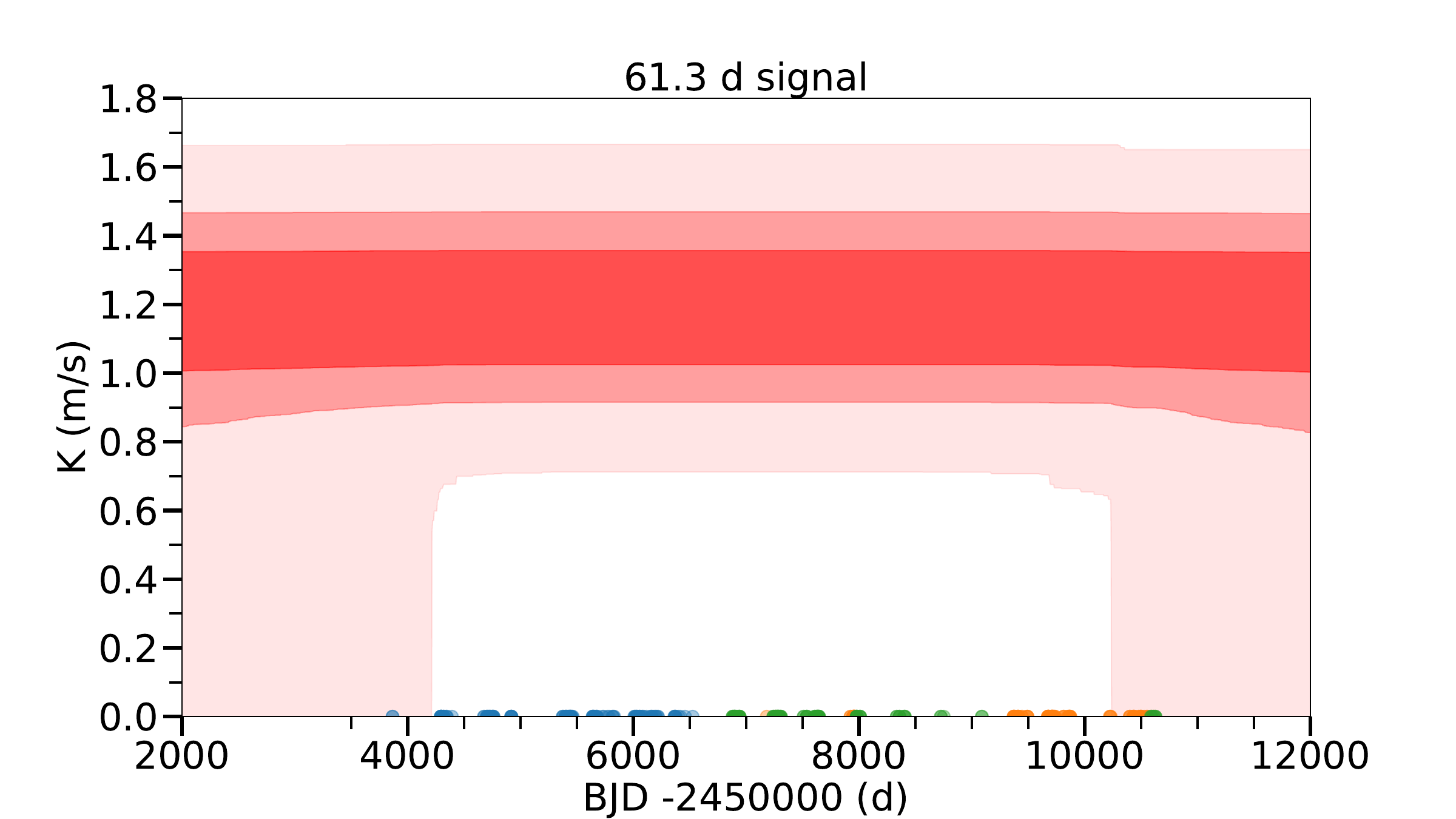}};
            
            \node[anchor=north west, inner sep=0] at (0.3, -0.6) {(c)}; 
        \end{tikzpicture}
        \label{fig:upper}
    \end{subfigure}
    \caption{Box test for the three planets orbiting around HD 176986. Color-coded dots on the x axis represents the observing epochs for each instrument. Panel (a): Box test for HD 176986 b. The planetary signal is stable at most of the observing epochs, with a decrease in amplitude after BJD = 2460000. Panel (b): Box test for HD 176986 c. We see how the planetary signal is stable at all the epochs of observation.
    Panel (c): Box test for HD 176986 d. The planetary signal is stable in amplitude at every epoch.}
    \label{fig_box}
\end{figure}
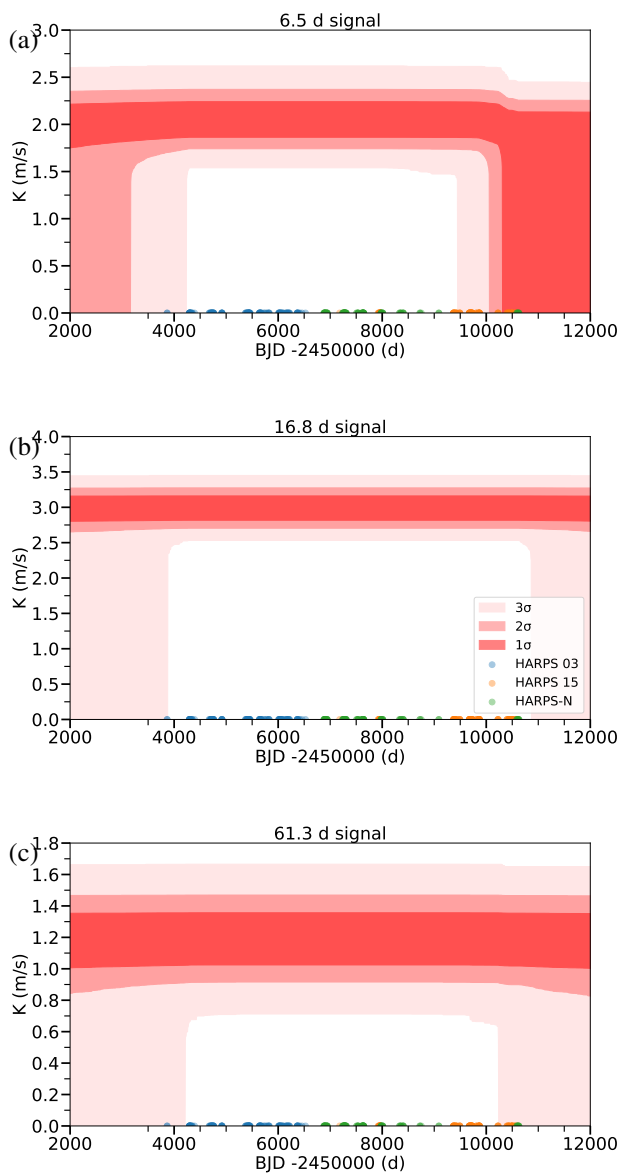

\section{Target for future atmospheric characterization}
\label{sec_atmospheric_charact}
Future missions, both from space and the ground, will be able to directly analyze the atmosphere of exoplanets. 

ANDES for the ELT \citep{2022_marconi_andes,2023_palle_andes} will exploit the capability of the 39-m telescope to resolve the angular distance between planets and their star. Disentangling the light coming from the star and the planet will make it possible to analyze the atmosphere of exoplanets through high-dispersion coronography (HDC). The angular resolution of ANDES at 1300 \si{\nano\meter} will be 13.7 mas \citep{2023_suarez_gj1002}. 
The angular separation of the three confirmed planets of HD 176986 is $\sim$ 2.3 mas for HD 176986 b, $\sim$ 4.3 mas for HD 176986 c, and 10.1 mas for HD 176986 d. The two inner planets are out of the technical limits of the instrument, while the third planet is close to the edge of the instrumental capability. Furthermore, the observational efficiency of a target depends on its planet-to-star contrast.  To calculate the planet-to-star contrast, we need an estimate of the minimum radius of the planets. We followed \citet{2020_otegi_mass_radius} to estimate the planetary radius. For the two smaller planets, we considered both the rocky and volatile-rich scenario, while for the more massive planet we only considered the volatile-rich scenario. The radius of HD 176986 b is $\sim$ 1.7 R$\oplus$ in the rocky planet scenario and $\sim$ 2.1 R$\oplus$ in the volatile-rich scenario. This implies a minimum planet-to-star contrast of $\sim$ $1.3 \times 10^{-7}$ in the rocky planet scenario and a contrast of $\sim$ $1.9 \times 10^{-7}$ for the volatile-rich scenario. The radius of HD 176986 c is estimated to be $\sim$ 3.0 R$\oplus$, which transform in a minimum planet-to-star contrast of $\sim$ $1.1 \times 10^{-7}$. The radius of HD 176986 d is estimated to be 1.8 R$\oplus$ in the rocky planet scenario and 2.4 in the volatile-rich scenario, with a minimum contrast of $\sim$ $7.3 \times 10^{-9}$ and $\sim$ $1.3 \times 10^{-8}$, respectively. We considered for our calculation an albedo of 0.3. ANDES will be sensitive to planets with a planet-to-star-contrast $>$ 10$^{-9}$. A planet in the inner conservative limit of the HZ (a = 0.591 $\pm$ 0.024 AU) needs to have a radius of $\sim$ 1.5 R$\oplus$. For a rocky planet, this would mean a mass of less than 4 M$\oplus$. With the detection limits derived in Sect. \ref{det_lim}, we cannot exclude the presence of rocky planets in the HZ observable with ANDES. 
Other missions are going to test a similar approach from space. The Habitable World Observatory (HWO) is a NASA mission that is scheduled to launch in the 2040s \citep{2024_mamajek_hwo}. HWO will be sensitive to a planet-to-star contrast of 10$^{-11}$. HWO will be able to characterize planets orbiting their parent star at an angular separation larger than 60-70 mas. This limitation excludes the possibility of analyzing the planets orbiting HD 176986 with HWO.  In Fig. \ref{hd176986_planetary_system}, we can see the optimistic HZ has an angular separation from the star of $\sim$ 40 mas, out of the possibilities of HWO. 

Another mission which is going to analyze the atmosphere of exoplanets from space is the Large Interferometer For Exoplanets (LIFE; \citealt{2022_life_quanz}. LIFE will use interferometry to analyze the mid-infrared light coming from the thermal emission of exoplanets instead of reflected light. This mission will be able to explore the presence of biosignatures such as ozone, methane, nitrous oxide, and others. 

\section{Planetary parameters}

We show in Table \ref{table_planets} the main parameters of the planets found in our analysis. We present in Table \ref{table_priors} the priors used in our best model, along with the inferred posterior distributions of the parameters. 
We used Eq. \ref{temp_equation} to calculate the equilibrium temperature.
\begin{table}[h!]
  \centering
  \begin{threeparttable} 
  \caption{Parameters for the planetary system of HD 176986. }
    \label{table_planets}
    \begin{tabular}{l r r r}
    \hline
    \hline
    \noalign{\smallskip}
    Parameter & HD 176986 b & HD 176986 c & HD 176986 d \\
    \noalign{\smallskip}
    \hline
    \noalign{\smallskip}
      T0 (BJD)  & 2460624.06 $\pm$ 0.17 & 2460614.69$^{+0.32}_{-0.31}$ & 2460614.69$^{+3.0}_{-3.1}$ \\
      P (d) & 6.49164$^{+0.00030}_{-0.00029}$ & 16.8124 $\pm$ 0.0015 & 61.376$^{+0.051}_{-0.049}$ \\
      K (\si{\meter\per\second}) & 2.14 $\pm$ 0.17 & 2.84 $\pm$ 0.18 & 1.28 $\pm$ 0.17 \\
      M$_p$$\sin$i (M$\oplus$) & 5.36 $\pm$ 0.44 & 9.75$^{+0.65}_{-0.64}$ &  6.76$^{+0.91}_{-0.92}$\\
      a (AU) & 0.062956 $\pm$ 0.00053 &  0.118730 $\pm$ 0.0010 & 0.28149 $\pm$ 0.0024 \\
      S (S$\oplus$) & 81.7$^{+9.3}_{-8.6}$ & 22.9$^{+2.6}_{-2.4}$ & 4.09$^{+0.47}_{-0.43}$ \\
      T$_{eq}$ (K) & 767 $\pm$ 21 & 558 $\pm$ 15 & 363 $\pm$ 10 \\
      \bottomrule
    \end{tabular}
    \tablefoot{The T0 here reported is the time of the inferior conjunction.}
    \medskip
    
    \begin{minipage}{0.5\textwidth}
        \raggedright
        
    \end{minipage}
  \end{threeparttable}
\end{table}

\begin{table*}[h!]
  \centering
  \begin{threeparttable}
    \caption{Priors and posteriors of our favorite model.} 
    \label{table_priors}
    \begin{tabular}{l r r r}
    \hline
    \hline
    \noalign{\smallskip}
    Parameter & Unit & Prior & Posterior \\
    \noalign{\smallskip}
    \hline
    \noalign{\smallskip}
      K$_b$  & \si{\meter\per\second} & $\mathcal{U}(0,3\sigma RV)$ & 2.14 $\pm$ 0.17 \\
      P$_b$ & d & $\mathcal{U}(5,10)$ & 6.49164 $_{-0.00029}^{+0.00030}$ \\
      Phase $_b$ &  & $\mathcal{U}(0,1)$ & 0.4998$_{-0.026}^{+0.027}$ \\
      K$_c$  & \si{\meter\per\second} & $\mathcal{U}(0,3\sigma RV)$ & 2.84 $\pm$ 0.18 \\
      P$_c$ & d & $\mathcal{U}(10,20)$ & 16.8124 $\pm$ 0.0015 \\
      Phase $_c$ & & $\mathcal{U}(0,1)$ & 0.750$_{-0.019}^{+0.018}$ \\
      K$_d$  & \si{\meter\per\second} & $\mathcal{U}(0,3\sigma RV)$ & 1.28 $\pm$ 0.17 \\
      P$_d$ & d & $\mathcal{U}(40,60)$ & 61.376$_{-0.49}^{+0.051}$ \\
      Phase $_d$ &  & $\mathcal{U}(0,1)$ & -0.2148$_{-0.049}^{+0.050}$ \\
      FWHM offset H03 & \si{\meter\per\second} & $\mathcal{N}(0,3\sigma FWHM)$& 1.6 $\pm$ 2.4 \\
      FWHM offset H15 & \si{\meter\per\second} & $\mathcal{N}(0,3\sigma FWHM)$ & -3.2 $\pm$ 2.5 \\
      FWHM offset HN & \si{\meter\per\second} & $\mathcal{N}(0,3\sigma FWHM)$ & -3.7$_{-2.1}^{+2.2}$\\
      FWHM log jitter H03 & & $\mathcal{U}(-5,5)$ & 1.16 $_{-0.11}^{+0.10}$\\ \\
      FWHM log jitter H15 & & $\mathcal{U}(-5,5)$ & 1.39 $\pm$ 0.10 \\
      FWHM log jitter HN & & $\mathcal{U}(-5,5)$ &  1.31 1.16 $_{-0.091}^{+0.089}$\\
      FWHM acc & \si{\meter\per\second\per\day} & $\mathcal{U}(-0.06,0.06)$ & -0.00019 $\pm$ 0.00055\\
      FWHM H03 drift & \si{\meter\per\second\per\day} & $\mathcal{U}(-0.03,0.03)$& -0.00194 $_{-0.00081}^{+0.00080}$ \\
      S-$_{MW}$x100 offset H03 &  & $\mathcal{N}(0,3\sigma S_{MW})$ & 0.34 $\pm$ 0.39\\
      S-$_{MW}$x100 offset H15 &  & $\mathcal{N}(0,3\sigma S_{MW})$ & -0.82 $\pm$ 0.43\\
      S-$_{MW}$x100 offset HN &  & $\mathcal{N}(0,3\sigma S_{MW})$ & -0.83$_{-0.38}^{+0.39}$\\
      S-$_{MW}$x100 log jitter H03 & & $\mathcal{U}(-10,5)$ & -0.93$_{-0.14}^{+0.13}$ \\
      S-$_{MW}$x100 log jitter H15 & & $\mathcal{U}(-10,5)$ & 0.23$_{-0.12}^{+0.11}$\\
      S-$_{MW}$x100 log jitter HN & & $\mathcal{U}(-10,5)$ & -0.44 $\pm$ 0.10\\
      RV offset H03 & \si{\meter\per\second} & $\mathcal{N}(0,3\sigma RV)$ & 0.21 $_{-0.51}^{+0.50}$ \\
      RV offset H15 & \si{\meter\per\second} & $\mathcal{N}(0,3\sigma RV)$ & 1.23 $_{-0.56}^{+0.55}$ \\
      RV offset HN & \si{\meter\per\second} & $\mathcal{N}(0,3\sigma RV)$ & 0.42 $_{-0.36}^{+0.37}$ \\
      RV log jitter H03 & & $\mathcal{U}(-5,5)$& 0.860$_{-0.078}^{+0.079}$\\
      RV log jitter H15 & & $\mathcal{U}(-5,5)$ & 0.607$_{-0.090}^{+0.094}$\\
      RV log jitter HN & & $\mathcal{U}(-5,5)$ & 0.590$_{-0.090}^{+0.089}$\\
      RV acc & \si{\meter\per\second\per\day} & $\mathcal{U}(-0.03,0.03)$ & -0.00017 $\pm$ 0.00018\\
      K$_{cycle}$ FWHM & \si{\meter\per\second} & $\mathcal{U}(0,3\sigma FWHM)$ & 10.0 $_{-1.7}^{+1.9}$\\
      P$_{cycle}$ & \si{\day} & $\mathcal{U}(2000,3000)$ &  2432$_{-59}^{+64}$\\
      Phase $_{cycle}$ & & $\mathcal{U}(0.4,1.4)$ & 1.136$_{-0.031}^{+0.034}$\\
      K$_{cycle}$ S$_{MW}$x100 & & $\mathcal{U}(0,3\sigma S_{MW})$& 2.21$_{-0.32}^{+0.35}$\\
      K$_{cycle}$ RV & \si{\meter\per\second} & $\mathcal{U}(0,3\sigma RV)$& 0.39 $_{-0.25}^{+0.32}$ \\
      Log A1 FWHM & & $\mathcal{U}(-10,5)$& 2.276$_{-0.084}^{+0.093}$\\
      A1 S$_{MW}$ & & $\mathcal{U}(-1,1)$ & 0.1896$_{-0.0076}^{+0.0079}$\\
      A1 RV & \si{\meter\per\second} & $\mathcal{U}(-1,1)$& 0.1372$_{-0.017}^{+0.016}$\\
      A2 RV & \si{\meter} & $\mathcal{U}(-20,20)$& 0.52$_{-0.10}^{+0.11}$\\
      A2 FWHM & \si{\meter} & $\mathcal{U}(-40,40)$& 0.41$\pm$ 0.21\\
      P$_{rot}$ & \si{\day} & $\mathcal{U}(25,45)$& 36.05$_{-0.71}^{+0.66}$\\
      Log Timescale & & $\mathcal{U}(3.5,7)$& 4.36 $_{-0.27}^{+0.26}$\\
      Log $\eta$ & & $\mathcal{U}(-5,5)$& -0.85 $\pm$ 0.14\\
      \bottomrule
    \end{tabular}
    \medskip 
    
    \begin{minipage}{0.5\textwidth}
        \raggedright
        .
    \end{minipage}
  \end{threeparttable}
\end{table*}
\section{Literature comparison}

\begin{figure*}[!h]
    \begin{minipage}{0.69\textwidth}
        \includegraphics[width=\linewidth]{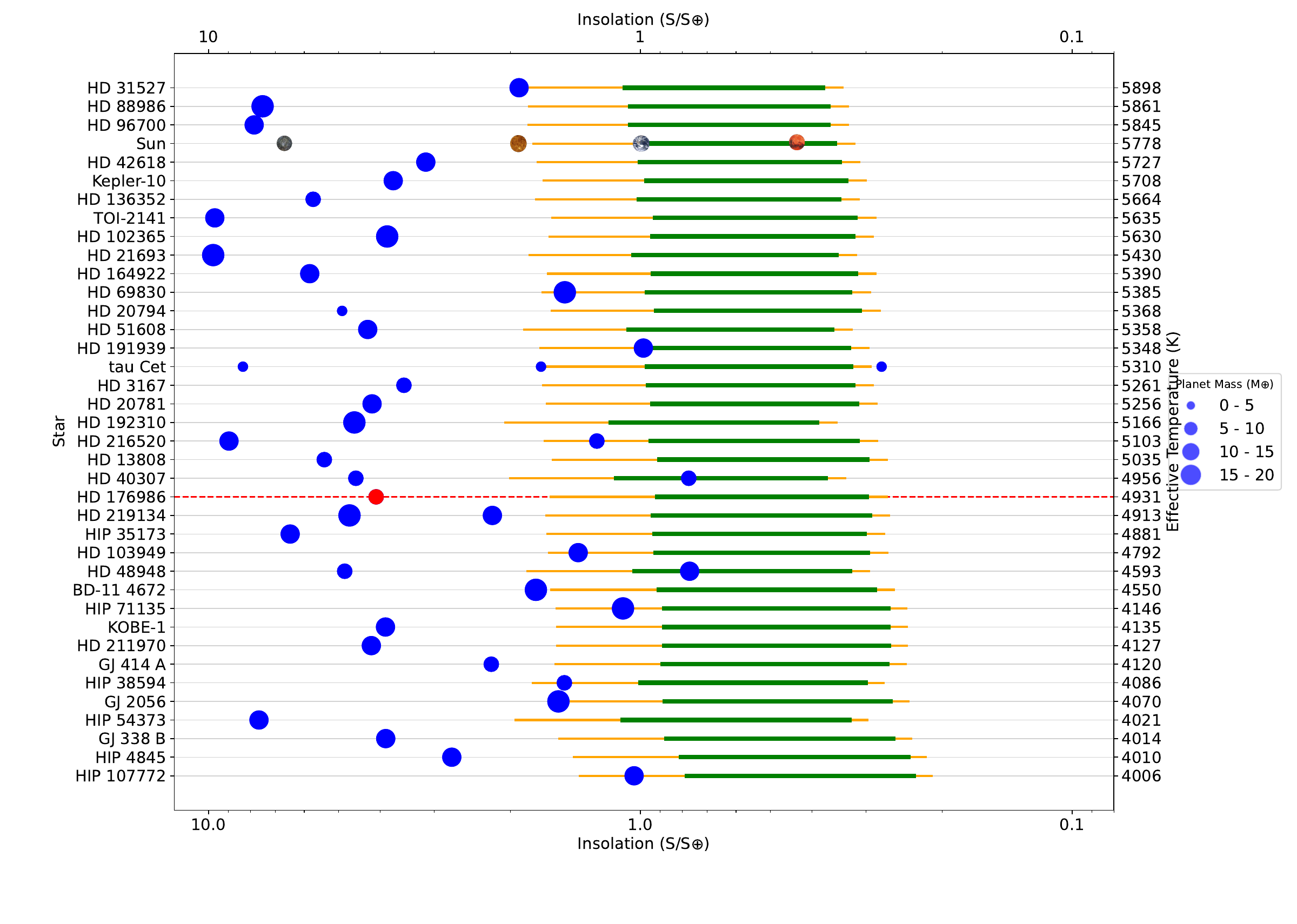}
    \end{minipage}
    \caption{Comparison between HD 176986 and other exoplanetary systems. Only three exoplanets have been discovered orbiting the conservative HZ of their parent star (green region), and 13 planets have been discovered orbiting the optimistic HZ (yellow region). Ephemeries of presented planets are taken from the NASA Exoplanet Archive \citep{2025_christiansen_archive}. 
    }
    \label{fig_comparison}
\end{figure*}

\section{Multidimensional GP}

\begin{figure*}[!h]
    \begin{minipage}{\textwidth}
        \includegraphics[width=\linewidth]{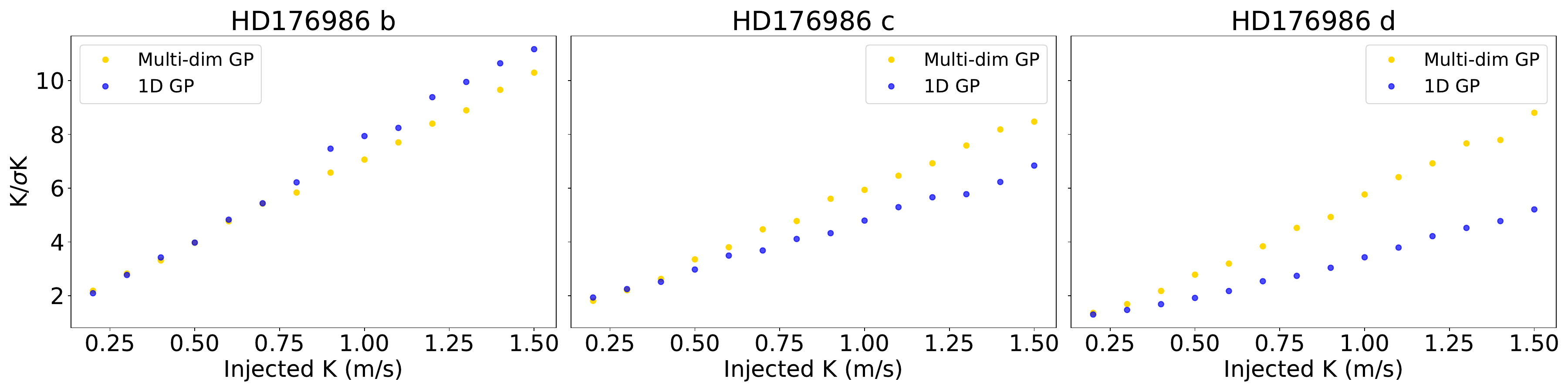}
    \end{minipage}
    
    \caption{Comparison of the $\sigma$ level detection of planetary injected signals at the periods of detected planets. 
    Panel (a): Injection-recovery of HD 176986 b. The 1D GP is slightly better. The orbital period of the planet is lower than half the rotation period.
    In this regime, the tendency of the 1D GP model to overfit is less pronounced.
    Panel (b): Injection-recovery of HD 176986 c. The orbital period of the planet is close to half the rotation period. At this period, the multidimensional GP outperforms the 1D GP.
    Panel (c): Injection-recovery of HD 176986 d. At a longer orbital period, the multidimensional GP framework outperforms the 1D GP.}
    \label{multi_dim_gp_overfitting}
\end{figure*}

\end{appendix}
\end{document}